\documentclass[seceq,preprint]{ptptex}

\usepackage{graphicx}
\usepackage{bm}

\preprintnumber[3cm]{
OCU-PHYS-240\\AP-GR-30\\PTPTeX ver.0.8\\ \today}

\title{
          Relativistic Gravitational Collapse of a Cylindrical Shell of 
Dust}

\author{
$^1$Ken-ichi \textsc{Nakao}, 
$^1$Yasunari \textsc{Kurita}, 
$^2$Yoshiyuki \textsc{Morisawa} \\
and \\
$^3$Tomohiro \textsc{Harada} 
}

\inst{

$^1$Department of Mathematics and Physics,~Graduate School of Science, \\
Osaka City University, Osaka 558-8585,~Japan \\
$^2$Faculty of Liberal Arts and Sciences, 
Osaka University of Economics and Law, \\
Yao City, Osaka 581-8511,~Japan \\
$^3$Department of Physics, Rikkyo University, Tokyo 171-8501,~Japan
}

\date{\today}

\abst{
The gravitational collapse of a thick cylindrical shell of dust matter 
is investigated. It is found that a spacetime singularity forms on 
the symmetry axis and that it is necessarily naked, i.e., observable in principle. 
We propose a physically reasonable boundary condition at this naked singularity to 
construct the solution including its causal future.
This boundary condition enables us to 
construct the unique continuation of spacetime beyond the 
naked singularity and ensures that the dust shell passes through 
the naked singularity. 
When the cylindrical shell leaves its symmetry axis away, the naked
singularity disappears, and regularity is recovered. 
We construct numerical solutions with this feature.
This result implies that the gravity produced by a thick cylindrical shell of 
dust is too weak to bind the shell even if it engenders the formation of 
a curvature singularity which is so strong as to 
satisfy the limiting focusing condition. 
For this reason, this naked singularity is very weak 
in the extended spacetime; the metric tensor is $C^{1-}$ even at the naked 
singularity, and the extended spacetime is complete for almost all geodesics.
This feature is also seen for singular hypersurfaces.
Such an extended spacetime can be regarded as phenomenological 
in the sense that it is valid if the relevant microphysics length scale 
is sufficiently small compared to the scale of interest.

}

\begin{document}

\maketitle

\section{Introduction}

General relativity predicts the gravitational collapse of very massive objects 
and the formation of spacetime singularities which will be accompanied by the 
divergence of physical quantities, such as the energy density, stress and spacetime 
curvature.\cite{Hawking-Ellis} 
An important issue related to the attribute of spacetime singularities is 
known as the cosmic censorship hypothesis \cite{Penrose:1969}. 
Roughly speaking, this hypothesis asserts the impossibility 
of the formation of an observable spacetime singularity, called 
a naked singularity in our universe. 
However, preceding studies have revealed that there are 
many possible counterexamples to this 
hypothesis. \cite{Eardley_S:1979,Chris:1984,Newman:1986,
Joshi_D:1993,Ori_P:1990,Joshi_K:1996,HIN:1998,Harada:1998,
Nakamura:1982a,Nakamura:1982b,Shapiro:1991} 

Although the appearance of spacetime singularities is an important 
consequence of general relativity, spacetime singularities 
themselves cannot be described in a physically reliable manner 
within the framework of general relativity, 
because general relativity becomes invalid 
beyond the quantum gravity scale. 
Even in the neighborhood of spacetime singularities, quantum 
effects in gravity should be important and in this sense, 
the neighborhood of singularities is also 
regarded as beyond the applicability of general 
relativity. In light of this observation, 
two of the present authors introduced the 
concept of a {\it spacetime border}. \cite{Harada-N:2004} 
Roughly speaking, a spacetime border 
is the spacetime region in which general relativity 
is not applicable due to the high curvature of spacetime within it. 
The implication of a possible counterexample to the cosmic 
censorship hypothesis would be that we can directly observe physical 
processes at the spacetime borders in our universe, 
other than the Big Bang.

Some of the known analytic solutions of the Einstein equations 
for naked singular spacetimes have been extensively investigated, 
and the structures and strengths of the corresponding naked singularities  
have been revealed. \cite{Eardley_S:1979,Chris:1984,Newman:1986,
Joshi_D:1993,Ori_P:1990,Joshi_K:1996,HIN:1998,Joshi:1993,KHI:2000,
Mena_N:2001,Nolan_M:2002,Deshingkar:2002,Nakao_KI:2003,Letelier:1994,
Nolan:2002,Kurita_N:2006} To determine whether 
a naked singularity is null or timelike and how a naked singularity 
is observed (such as in gravitational redshift), we have to take the limit
to a naked singularity from its causal future. 
This is an inevitable dilemma. We want to predict the observational
signature of naked singularities, while naked singularities 
are what prevent us from constructing the unique time development of the 
spacetime beyond them. In order to construct a solution 
of the Einstein equations 
for the causal future of a naked singularity, the boundary condition 
at the naked singularity must be specified, because spacetime 
singularities are boundaries of the spacetime. 

It might be thought of that a unique boundary condition at naked 
singularities cannot be selected from a large, maybe 
even infinite number of possible conditions, 
because we do not understand the laws of physics which describe physical 
processes in the spacetime border covering a naked singularity. 
However, we should note that for all of the known naked solutions, some 
boundary conditions at naked singularities have been implicitly imposed
without understanding the physical laws applicable to spacetime borders. 
One method for obtaining solutions for the causal future 
of naked singularities formed from regular initial data 
is to assume the extension across the Cauchy horizon associated with them
to be as smooth as possible. A typical example is the subcritical 
Reissner-Nordstr\"{o}m solution. Many examples 
of naked singular spacetimes are in this category, in which
analyticity on the Cauchy horizon is required. 
In other words, many known naked singular solutions 
are obtained by imposing boundary conditions which make 
possible smoothness on its Cauchy horizon, although nobody yet knows 
whether this is valid from the point of view of a quantum 
theory of gravity. It is important to elucidate the 
principle regarding the determination of boundary conditions at 
naked singularities in known analytic solutions, but this is 
beyond the scope of this paper. 

An exceptional example of naked singularities is the 
class of singular hypersurfaces 
on which some scalar polynomials of the curvature tensor diverge. 
It is believed that they can be described in a reliable manner
with the junction condition~\cite{Israel}.
In fact, it seems that many people use a singular hypersurface 
without realizing that this introduces a curvature singularity. 
It is not difficult to understand why this is possible.
If the hypersurface is actually infinitesimally thin, of course
the energy density diverges, and therefore 
it corresponds to a curvature singularity.
However, to derive the junction condition, we first assume
a finite thickness $\epsilon$, integrate the Einstein equation 
along the normal direction, and then take the limit $\epsilon\to 0$, 
keeping the surface density and pressure constant.
After taking this limit, the metric tensor can still be 
locally $C^{1-}$, in which case geodesics are uniquely extended across 
the singular hypersurface~\cite{Hawking-Ellis}.
A singular hypersurface is regarding as describing the 
gravitational effects of a hypersurface of very 
small but finite thickness.
Hence, a singular hypersurface can be regarded as providing 
a phenomenological description of a physical
hypersurface of small but finite thickness. 
This is also the reason why the singular hypersurface formalism should 
be capable of treating gravitational effects of a hypersuface whose 
thickness is as small as the quantum gravity scale, such as a 
$D$-brane.\cite{D-brane} 
To proceed with this approach, if we can find a $C^{1-}$
extension of spacetime beyond a naked singularity, 
it could be regarded as a phenomenological description of 
something whose length scale is much smaller than the scale of interest, 
because of the unique extendibility of geodesics across singularities.

In this paper, we investigate the gravitational collapse of a 
cylindrically symmetric thick dust shell surrounding a vacuum 
interior, in other words, a hollow cylinder composed 
of dust matter. Such a configuration of matter causes the formation of 
a string-like naked singularity 
of codimension two along the symmetry axis.\cite{Thorne:1972,Hayward:2000} 
We propose a physically reasonable boundary condition 
at this naked singularity. 
It is difficult to analytically obtain solutions for this system.
Therefore we numerically solve the basic equations. 
The obtained solution is a $C^{1-}$ extension of spacetime, and 
geodesics uniquely extend across the naked singularity, 
except for that which is on the string-like naked singularity. 
This system is not asymptotically flat, and thus it alone 
does not constitute a counterexample to weak cosmic censorship. 
However, this model might describe a portion of the spindle 
collapse studied by Shapiro and Teukolsky, 
which is an interesting possibility as a counterexample to the cosmic 
censorship hypothesis. \cite{Shapiro:1991} 

This paper is organized as follows. In $\S$2, we give the basic 
equations for the gravitational collapse of dust matter in a 
spacetime with whole cylinder symmetry. 
In $\S$3, we propose physically reasonable 
boundary conditions on metric and matter variables 
on the symmetry axis where a naked singularity is caused to form by 
the cylindrical gravitational collapse of a thick dust shell. 
Then, we discuss the physical meaning of such boundary conditions. 
In $\S$4, we present a numerical solution that describes the gravitational 
collapse of a thick dust shell. 
In $\S$5, we show that almost all of the causal geodesics are complete 
in the spacetime considered in this paper.  
In $\S$6, the strength of the curvature at the 
naked singularity is investigated. Finally, $\S$7 is devoted to summary 
and discussion. In Appendices A and B, we prove two facts which are very 
important in in the investigation carried out in this paper. In Appendix C,  
the Christoffel symbols of the spacetime with whole cylinder symmetry 
in the Cartesian coordinate system are given, and 
in Appendix D, the components of the Riemann tensor in the cylindrical 
coordinate system are given. 

In this paper, we adopt units in which $c=1$. Greek indices denote  
components with respect to the coordinate basis. 

\section{Basic equations for a cylindrical dust system}

We consider the spacetime with whole cylinder symmetry \cite{Melvin:1964,Melvin:1965} 
defined by the metric 
\begin{equation}
ds^{2}=e^{2(\gamma-\psi)}\left(-dt^{2}+dr^{2}\right)
+r^2\beta^2e^{-2\psi}d\varphi^{2}+e^{2\psi}dz^{2},  \label{eq:cylindrical}
\end{equation}
where $0\leq r<+\infty$, $0\leq\varphi<2\pi$ and 
$-\infty<z<+\infty$ constitute the cylindrical coordinate system, 
and $\gamma$, $\psi$ and $\beta$ are functions of the time coordinate $t$ and 
the radial coordinate $r$. The coordinate basis vectors $\partial/\partial\varphi$ 
and $\partial/\partial z$ are Killing vectors. 
The coordinates $t$, $r$ and $z$ are all normalized so as 
to be dimensionless. 

We consider a two-component model of dust matter and assume that 
one component is collapsing while the other component is expanding and 
that these components do not interact with each other. 
The stress-energy tensors of the collapsing dust, $T^{\mu\nu}_{\rm i}$, and 
the expanding dust, $T^{\mu\nu}_{\rm o}$, are, respectively, 
written in the forms 
\begin{equation}
T_{\rm i}^{\mu\nu}=\rho u^\mu u^\nu,~~~~~
T_{\rm o}^{\mu\nu}=\mu v^\mu v^\nu,
\end{equation}
where $\rho$ and $\mu$ are the rest mass densities, whereas 
$u^\mu$ and $v^\mu$ are the 4-velocities of the 
collapsing and expanding components of the dust matter, respectively. 
We assume that $\rho$ and $\mu$ are non-negative so that 
all of the energy conditions are satisfied. 
By virtue of the assumed symmetry of the spacetime, 
the 4-velocities $u^\mu$ and $v^\mu$ can be written in the forms 
\begin{equation}
u^\mu=\frac{e^{-\gamma+\psi}}{\sqrt{1-u^2}}\left(1,u,0,0\right),~~~~
v^\mu=\frac{e^{-\gamma+\psi}}{\sqrt{1-v^2}}\left(1,v,0,0\right),
\end{equation}
where $u~(\leq0)$ and $v~(\geq0)$ correspond to the 3-velocities of the collapsing 
and expanding components of the dust matter, respectively. 

In place of the rest mass densities $\rho$ and $\mu$, 
we introduce the conserved rest mass densities $D$ and $E$ defined by
\begin{eqnarray}
D&:=&\sqrt{-g}\rho u^t=\frac{r\beta e^{\gamma-\psi}\rho}{\sqrt{1-u^2}}, \label{eq:D-def} \\
E&:=&\sqrt{-g}\mu v^t=\frac{r\beta e^{\gamma-\psi}\mu}{\sqrt{1-v^2}}, \label{eq:E-def}
\end{eqnarray}
where $g$ is the determinant of the metric tensor. 
Then the equations of motion for each component of the dust matter, 
$\nabla_\nu T_{\rm o}^{\mu\nu}=0$ and $\nabla_\nu T_{\rm i}^{\mu\nu}=0$, 
become
\begin{eqnarray}
\dot{D}+\left(uD\right)'&=&0, \label{eq:D-evol}\\
\dot{E}+\left(vE\right)'&=&0, \label{eq:E-evol}\\
{du\over dt}=\dot{u}+uu'&=&(1-u^2)\left[u\left(\dot{\psi}-\dot{\gamma}\right)
+\psi'-\gamma'\right],
\label{eq:U-evol} \\
{dv\over dt}=\dot{v}+vv'&=&(1-v^2)\left[v\left(\dot{\psi}-\dot{\gamma}\right)
+\psi'-\gamma'\right],
\label{eq:V-evol}
\end{eqnarray}
where the dot represents the time derivative, while the dash represents the radial derivative. It should be noted that the first and second equations represent 
conservation of the rest mass, and for this reason 
we refer to $D$ and $E$ as the conserved rest mass densities, 
whereas Eqs.~(\ref{eq:U-evol}) and (\ref{eq:V-evol}) are 
the geodesic equations for the constituent 
particles of the dust matter. 

The Einstein equations take the forms 
\begin{eqnarray}
&&\gamma'=
\left\{\beta^2+2r\beta\beta'+r^2({\beta'}^{2}-{\dot \beta}^{2})\right\}^{-1}
\biggl[
r\beta\left(\beta+r\beta'\right)\left({\dot \psi}^{2}+{\psi'}^{2}\right)
-2r^2\beta\dot{\beta}{\dot \psi}\psi' \nonumber \\
&&~~~~+2\beta\beta'+r\left(2{\beta'}^2+\beta\beta''-\dot{\beta}^2\right)
+r^2\left(\beta'\beta''-\dot{\beta}\dot{\beta}'\right)\nonumber \\ 
&&~~~~+\frac{8\pi Ge^{\gamma-\psi}D}{\sqrt{1-u^2}}
 \left(\beta+r\beta'+r\dot{\beta}u\right)
+\frac{8\pi Ge^{\gamma-\psi}E}{\sqrt{1-v^2}}
 \left(\beta+r\beta'+r\dot{\beta}v\right)
\biggr], 
\label{eq:constraint-1} \\
&&{\dot \gamma}=-
\left\{\beta^2+2r\beta\beta'+r^2({\beta'}^{2}-{\dot \beta}^{2})\right\}^{-1}
\biggl[
r^2\beta\dot{\beta}\left({\dot \psi}^{2}+{\psi'}^{2}\right)
-2r\beta\left(\beta+r\beta'\right){\dot \psi}\psi' \nonumber \\
&&~~~~-\beta\dot{\beta}+r\left(\dot{\beta}\beta'-\beta\dot{\beta}'\right)
+r^2\left(\dot{\beta}\beta''-\beta'\dot{\beta}'\right)\nonumber \\
&&~~~~+\frac{8\pi Ge^{\gamma-\psi}D}{\sqrt{1-u^2}}
\left\{r\dot{\beta}+(\beta+r\beta')u\right\}
+\frac{8\pi Ge^{\gamma-\psi}D}{\sqrt{1-v^2}}
\left\{r\dot{\beta}+(\beta+r\beta')v\right\}
\biggr], 
\label{eq:constraint-2}\\
&&{\ddot \beta}-\beta''-{2\over r}\beta'
={8\pi G\over r}e^{\gamma-\psi}
\left(D\sqrt{1-u^2}+E\sqrt{1-v^2}\right), \label{eq:beta-evol}\\
&&{\ddot \gamma}-\gamma''={\psi'}^{2}-{\dot \psi}^{2}, \label{eq:gamma-evol}\\
&&{\ddot \psi}+{{\dot \beta}\over \beta}{\dot \psi}-\psi''
-{1\over r}\left(1+r{\beta'\over \beta}\right)\psi' 
={4\pi G\over r\beta}e^{\gamma-\psi} \nonumber \\
&&~~~~~~~~~~~~~~~~~~~~~~~~~~~~~~~~~~~~~~~~~~
\times\left(D\sqrt{1-u^2}+E\sqrt{1-v^2}\right).
\label{eq:psi-evol}
\end{eqnarray}
Equations (\ref{eq:constraint-1}) and (\ref{eq:constraint-2}) 
are the constraint equations, and 
Eqs.~(\ref{eq:beta-evol})--(\ref{eq:psi-evol}) 
are the evolution equations for the 
metric variables $\beta$, $\gamma$ and $\psi$. 

\section{ Boundary condition}

\subsection{Spacetime regularity}

First, we stress that it is possible to construct $C^{2-}$ solutions 
for the metric variables, $\beta$, $\gamma$ and $\psi$, and 
everywhere finite solutions for the matter variables, 
$D$, $E$, $u$ and $v$, as functions of $t$ and $r$, even if 
the spacetime singularity is caused to form 
at $r=0$ by the gravitational collapse of the cylindrical dust matter. 
This may seem strange. However, it 
should be noted that it is not a sufficient condition for 
the realization of regular spacetimes that the metric variables 
be $C^{2-}$ functions of $t$ and $r$. As shown below, 
in order for the spacetime to be regular, the first-order 
radial derivative of the metric variables $\beta'$, $\gamma'$ and 
$\psi'$ should vanish at $r=0$,  whereas 
the matter variables $D$, $E$, $u$ and $v$ 
should vanish at $r=0$. 
In order to understand the meaning of these conditions, 
we introduce a Cartesian coordinate system defined by 
\begin{equation}
x=r\cos\varphi, ~~~~~~~~
y=r\sin\varphi, \label{eq:Cartesian}
\end{equation}
with the remaining coordinates, $t$ and $z$, unchanged. 
In the case of the regular spacetime, the coordinate singularity 
at $r=0$ in the cylindrical coordinate system does not exist in this 
Cartesian coordinate system. 

We have the following components of the metric tensor 
in this new coordinate basis: 
\begin{eqnarray}
g_{tt}&=&-e^{2(\gamma-\psi)}, \label{eq:g_tt}\\
g_{xx}&=&\beta^2e^{-2\psi}
\left[\left(\frac{e^{2\gamma}}{\beta^2}-1\right)\cos^2\varphi+1\right], \\
g_{xy}&=&\beta^2e^{-2\psi}\left(\frac{e^{2\gamma}}{\beta^2}-1\right)
\sin\varphi\cos\varphi, \\
g_{yy}&=&\beta^2e^{-2\psi}
\left[\left(\frac{e^{2\gamma}}{\beta^2}-1\right)\sin^2\varphi+1\right],\\
g_{zz}&=&e^{2\psi}, \label{eq:g_zz}
\end{eqnarray}
with all other components vanishing. 
It can be easily seen from the above equations that 
$\beta^2$ should be equal to $e^{2\gamma}$ at $r=0$ 
in order for the components of the metric tensor to be 
single-valued at $r=0$. 
(Note that any value can be assigned to $\varphi$ at $r=0$.) 
The first and second order derivatives of the components of 
the metric tensor in the Cartesian coordinate system with respect to 
$x$ and $y$ are written in the form
\begin{eqnarray}
\frac{\partial g_{\mu\nu}}{\partial x}&=&g_{\mu\nu}'\cos\varphi
+\frac{1}{r}\frac{\partial g_{\mu\nu}}{\partial \varphi}
\sin\varphi,\\
\frac{\partial g_{\mu\nu}}{\partial y}&=&g_{\mu\nu}'\sin\varphi
-\frac{1}{r}\frac{\partial g_{\mu\nu}}{\partial \varphi}
\cos\varphi, \\
\frac{\partial^2 g_{\mu\nu}}{\partial x^2}&=&
\frac{1}{2}g_{\mu\nu}''(1+\cos2\varphi)
+\frac{1}{2r}\left[g_{\mu\nu}'(1-\cos2\varphi)
-\frac{\partial g_{\mu\nu}'}{\partial\varphi}
\sin2\varphi\right] \nonumber \\
&+&\frac{1}{r^2}\left[\frac{\partial g_{\mu\nu}}{\partial\varphi}
\sin2\varphi +\frac{1}{2}\frac{\partial^2 g_{\mu\nu}}{\partial\varphi^2}
(1-\cos2\varphi)
\right], \\
\frac{\partial^2 g_{\mu\nu}}{\partial x\partial y}&=&
\frac{1}{2}g_{\mu\nu}''\sin2\varphi
-\frac{1}{2r}\left[g_{\mu\nu}'\sin2\varphi
-\frac{\partial g_{\mu\nu}'}{\partial\varphi}
(1+\cos2\varphi)\right] \nonumber \\
&-&\frac{1}{r^2}\left[\frac{\partial g_{\mu\nu}}{\partial\varphi}
\cos2\varphi +\frac{1}{2}\frac{\partial^2 g_{\mu\nu}}{\partial\varphi^2}
\sin2\varphi
\right], \\
\frac{\partial^2 g_{\mu\nu}}{\partial y^2}&=&
\frac{1}{2}g_{\mu\nu}''(1-\cos2\varphi)
+\frac{1}{2r}\left[g_{\mu\nu}'(1+\cos2\varphi)
+\frac{\partial g_{\mu\nu}'}{\partial\varphi}
\sin2\varphi\right] \nonumber \\
&-&\frac{1}{r^2}\left[\frac{\partial g_{\mu\nu}}{\partial\varphi}
\sin2\varphi -\frac{1}{2}\frac{\partial^2 g_{\mu\nu}}{\partial\varphi^2}
(1+\cos2\varphi)
\right].
\end{eqnarray}
We can see from the above equations and Eqs.~(\ref{eq:g_tt})-(\ref{eq:g_zz}) 
that in order for the components of the metric tensor 
$g_{\mu\nu}$ in the Cartesian coordinate system to be 
$C^{2-}$ functions of $x$ and $y$, 
the following regularity conditions must be satisfied: 
\begin{eqnarray}
&&\lim_{r\rightarrow0}\frac{e^{2\gamma}-\beta^2}{r^2}=[{\rm finite}],
\label{eq:reg-metric-1} \\
&&\lim_{r\rightarrow0}\frac{\beta'}{r}=[{\rm finite}],~~
\lim_{r\rightarrow0}\frac{\gamma'}{r}=[{\rm finite}],~~
\lim_{r\rightarrow0}\frac{\psi'}{r}=[{\rm finite}], \label{eq:reg-metric-2} \\
&&\lim_{r\rightarrow0}\beta''=[{\rm finite}],~~
\lim_{r\rightarrow0}\gamma''=[{\rm finite}],~~
\lim_{r\rightarrow0}\psi''=[{\rm finite}]. \label{eq:reg-metric-3}
\end{eqnarray}
Note that the above regularity conditions 
do not guarantee the continuity of the second order derivatives 
of the metric components at the symmetry axis, $r=0$. Although this does 
not necessarily mean that the metric tensor is not $C^{2}$, it might imply 
that the Cartesian coordinate system (\ref{eq:Cartesian}) itself 
is $C^{2-}$. However, this Cartesian coordinate system is sufficient for our 
present purpose. 

We can see from Eqs.~(\ref{eq:D-def}) and (\ref{eq:E-def}) that 
$D$ and $E$ vanish on the symmetry axis, $r=0$, 
if the rest mass densities $\rho$ and $\mu$ and the metric variables are 
everywhere finite, and, further, if the absolute values of 
$u$ and $v$ are smaller than unity. 
The components of the 4-velocities $u^\mu$ and $v^\mu$ in the Cartesian coordinate system are given by 
\begin{eqnarray}
u^\mu&=&\frac{e^{-\gamma+\psi}}{\sqrt{1-u^2}}
\left(1,u\cos\varphi,u\sin\varphi,0\right), \\
v^\mu&=&\frac{e^{-\gamma+\psi}}{\sqrt{1-v^2}}
\left(1,v\cos\varphi,v\sin\varphi,0\right).
\end{eqnarray}
From the above equations, it is seen that 
in order for the components of $u^\mu$ and $v^\mu$ in the 
Cartesian coordinate system to be everywhere single-valued, 
$u$ and $v$ must vanish at $r=0$. Thus, the regularity 
condition on the matter variables are 
\begin{equation}
\lim_{r\rightarrow0}D=\lim_{r\rightarrow0}E=0~~~{\rm and}~~~
\lim_{r\rightarrow0}u=\lim_{r\rightarrow0}v=0. \label{eq:reg-matter}
\end{equation}

\subsection{Functional regularity and boundary conditions on metric components}

If a spacetime singularity is formed by the collapse of the cylindrical 
dust matter, the regularity conditions on the metric variables, 
(\ref{eq:reg-metric-1}) and (\ref{eq:reg-metric-2}), and on the 
matter variables, (\ref{eq:reg-matter}), will not be satisfied. 
From Eq.~(\ref{eq:beta-evol}), we have 
\begin{equation}
\beta'=-4\pi Ge^{\gamma-\psi}
\left(
D\sqrt{1-u^2}+E\sqrt{1-v^2}
\right)+r\left(\ddot{\beta}-\beta''\right).
\end{equation}
Because we construct $C^{2-}$ solutions for the metric variables and 
everywhere finite solutions for the matter variables, the above equation 
gives the following Neumann boundary condition on $\beta$ at $r=0$: 
\begin{equation}
\beta'|_{r=0}=-4\pi Ge^{\gamma-\psi}
\left(
D\sqrt{1-u^2}+E\sqrt{1-v^2}
\right)
\Bigl|_{r=0}~. 
\label{eq:beta-bc}
\end{equation}
With the same procedure, we obtain the following Neumann 
boundary conditions on $\psi$ and $\gamma$ at $r=0$ from 
Eqs.~(\ref{eq:constraint-1}) and (\ref{eq:psi-evol}): 
\begin{eqnarray}
\gamma'|_{r=0}&=&\frac{8\pi Ge^{\gamma-\psi}}{\beta}
\left(
\frac{Du^2}{\sqrt{1-u^2}}+\frac{Ev^2}{\sqrt{1-v^2}}
\right)
\biggr|_{r=0}~, \label{eq:gamma-bc} \\
\psi'|_{r=0}&=&-{4\pi G \over \beta}e^{\gamma-\psi}
\left(
D\sqrt{1-u^2}+E\sqrt{1-v^2}
\right)
\Bigr|_{r=0}~. \label{eq:psi-bc}
\end{eqnarray}
Here we have used Eq.~(\ref{eq:beta-bc}) to derive Eq.~(\ref{eq:gamma-bc}). 

Because it is difficult to analytically construct solutions 
for cylindrically symmetric spacetimes with dust matter,  
numerical simulations are necessary to study their detailed behavior. 
The boundary conditions (\ref{eq:beta-bc})--(\ref{eq:psi-bc}) 
are sufficient to construct numerical solutions 
for the evolution equations  (\ref{eq:beta-evol})--(\ref{eq:psi-evol}) 
without any ambiguities, if the matter variables 
$D$, $E$, $u$ and $v$ are everywhere finite. 

There will be three kinds of spacetime singularities formed through the 
collapse of a cylindrical dust matter. 
The first kind is the so-called {\it shell crossing singularity}, 
whose appearance is caused by a caustic of dust matter at 
some position not on the symmetry axis and is necessarily accompanied by 
the divergence of one or both of the conserved densities $D$ and $E$. 
However, it is known that shell crossing 
singularities are so weak that they are not physically so 
significant. For this reason, we do not consider shell crossing 
singularities here. The second kind of spacetime singularity 
forms on the symmetry axis, $r=0$, without 
a caustic of the dust matter. In this case, 
the matter variables $D$, $E$, $u$ and $v$ are everywhere finite, 
even if a spacetime singularity does exist. 
This is the case of interest in this paper. 
Since this kind of spacetime singularity is formed by  
the dust matter focused on $r=0$, it is often called a 
{\it shell focusing singularity}. 
The third kind of spacetime singularity is also a shell focusing singularity, 
but with a ``caustic'' of dust matter at $r=0$. In this case, $D$ or $E$ diverges at $r=0$, and thus it is impossible for the metric variables 
to remain $C^{2-}$ and the matter variables to remain 
everywhere finite, even if we impose the 
boundary conditions (\ref{eq:beta-bc})--(\ref{eq:psi-bc}). In this paper, we 
do not consider this kind of spacetime singularity in detail, 
but we briefly discuss it in $\S$7. If $\beta$ vanishes at some point, 
it would seem that another kind of spacetime singularity is 
formed there, as several terms in the Einstein equations 
(\ref{eq:constraint-1})--(\ref{eq:psi-evol}) 
are proportional to $\beta^{-1}$. However, 
as shown in Appendix A, this possibility is excluded by the boundary 
conditions on the matter variables. 

Hereafter, we focus on the second kind of spacetime singularity. 
Now we can see how a spacetime singularity forms on the symmetry 
axis $r=0$ through the gravitational collapse of the dust matter 
while the metric variables $\beta$, $\gamma$ and $\psi$  
remain $C^{2-}$ and the matter variables $D$, $E$, $u$ and $v$ 
remain everywhere finite. 
Since we are interested in gravitational collapse, we consider the 
initial data with the only collapsing dust $(D,u)$. 
To avoid the appearance of a caustic before the 
formation of the second kind of 
spacetime singularity, we assume that the 3-velocity field 
$u$ is initially a monotonically increasing function of $r$. 
(Here note that we have $u\leq0$, by definition.) 
If $\rho$ initially does not vanish on the symmetry axis $r=0$ with 
such an initial 3-velocity field, 
the regularity condition on the 3-velocity field $u$ 
in (\ref{eq:reg-matter}) cannot be satisfied. 
Thus, we have to assume that the initial configuration of dust matter 
is a cylindrical thick shell, i.e., a hollow cylinder. 
(The case of a solid cylinder is discussed in $\S$7.) 
When a collapsing thick dust shell reaches the symmetry axis, $r=0$, 
$D$ becomes non-vanishing at $r=0$. Then the rest mass density 
$\rho$ of the collapsing dust becomes infinite, because 
$\rho$ is proportional to $D/r$, as seen from Eq.~(\ref{eq:D-def}). 
Because the Ricci scalar is proportional to $\rho+\mu$ as given 
by the Einstein equations, the divergence of $\rho$ implies 
the formation of a scalar polynomial ({\it s.p.}) 
curvature singularity.\cite{Hawking-Ellis} It is also easily seen 
from Eqs.~(\ref{eq:beta-bc})--(\ref{eq:psi-bc}) that 
if $D$ does not vanish at $r=0$, then $\beta'$, 
$\gamma'$ and $\psi'$ do not vanish at $r=0$, or equivalently, 
the regularity of spacetime is broken. 

\subsection{Boundary conditions on dust matter}

It should be noted that we still have freedom in the choice of the boundary 
conditions on the matter variables $D$, $E$, $u$ and $v$ imposed at $r=0$. 
Here we impose the following boundary conditions on the matter variables 
at the symmetry axis, $r=0$:
\begin{equation}
E(t,0)=D(t,0)~~~{\rm and}~~~v(t,0)=-u(t,0).
\end{equation}
The above boundary conditions physically mean that once the collapsing dust 
reaches the symmetry axis $r=0$, the same amount of the expanding dust 
is shot from $r=0$ with the same speed as the collapsing dust matter. 

In order to impose the above boundary conditions in numerical simulations, 
it is convenient to extend the domain $0\leq r<\infty$ to 
$-\infty<r<\infty$; we call the original domain {\it the physical domain} 
and the additional domain, $-\infty<r<0$, {\it the fictitious domain}. 
We define the metric variables in the fictitious domain 
such that these variables possess reflection symmetry with 
respect to $r=0$, i.e., 
\begin{eqnarray}
\gamma(t,-r)=\gamma(t,r) ~~~{\rm and}~~~\psi(t,-r)=\psi(t,r). \label{eq:ref-condition}
\end{eqnarray}
Then we solve Eqs.~(\ref{eq:D-evol}) and (\ref{eq:U-evol}) for 
the variables of the collapsing dust $(D,u)$ in the fictitious 
domain, $r<0$, as well as in the physical domain, $r\geq0$. 
For the variables of the expanding dust $(E,v)$, we impose the conditions
\begin{equation}
E(t,r)=D(t,-r)~~~{\rm and}~~~v(t,r)=-u(t,-r). \label{eq:PT-condition}
\end{equation}
The variables $E$ and $v$ determined by the above condition automatically  
satisfy Eqs.~(\ref{eq:E-evol}) and (\ref{eq:V-evol}). 

Here it should be noted that in the extended domain, $-\infty <r<\infty$, 
the metric variables are no longer $C^{2-}$, even though they are regular 
with respect to $t$ and $r$ in the physical domain, $0\leq r<\infty$. 
If a spacetime singularity forms, $\gamma'$ and $\psi'$ do not vanish at 
the symmetry axis, $r=0$. Then the reflection symmetry condition 
(\ref{eq:ref-condition}) leads discontinuities of $\gamma'$ and $\psi'$ 
at $r=0$ in the extended domain. This low differentiability  
in the extended domain causes a decrease in the accuracy of numerical simulations, and thus we need a careful treatment to maintain good 
numerical accuracy. 

\subsection{Physical consequences of the  boundary conditions}

The above conditions on the matter variables lead the following 
three physically favorable consequences. 

\vskip0.3cm
\noindent
(i) $C^{1-}$ metric

\vskip0.3cm
It should be noted that the boundary condition (\ref{eq:PT-condition}) 
guarantees that the metric tensor is defined even at the spacetime singularity 
at $r=0$. From Eq.~(\ref{eq:constraint-2}), we find the relation 
\begin{equation}
\frac{\partial}{\partial t}\left(\frac{e^\gamma}{\beta}\right)=-8\pi Ge^{-\psi}
\left(\frac{e^\gamma}{\beta}\right)^2
\left(\frac{Du}{\sqrt{1-u^2}}+\frac{Ev}{\sqrt{1-v^2}}
\right)~~~~{\rm at}~~r=0
\end{equation}
is satisfied. Before the singularity formation, $e^{2\gamma}/\beta^2$ 
should be unity. (Hereafter, without loss of generality, 
we assume $e^\gamma/\beta=1$ at $r=0$ before the singularity formation.) 
Otherwise, the symmetry axis, $r=0$, would be conically singular 
even before the thick dust shell collapses to the symmetry axis, $r=0$. 
Thus the condition (\ref{eq:PT-condition}) and the above equation always guarantee the relation
\begin{equation}
\frac{e^\gamma}{\beta}=1~~~~~~{\rm at}~~r=0. \label{eq:no-conical}
\end{equation}
As previously shown, this implies that the metric tensor 
is finite and single-valued, even at the spacetime singularity, 
or, in other words, it is at least $C^0$. 
Further, it is not such a difficult task to show that the components of 
the metric tensor in the Cartesian coordinate system are $C^{1-}$, i.e., 
locally Lipschitz functions. 

\vskip0.3cm
\noindent
(ii) Completeness of radial geodesics

\vskip0.3cm
As shown in $\S$5, the radial geodesics that intersect 
the spacetime singularity at $r=0$ from the domain $r>0$ 
can be extended across the spacetime singularity. 
This implies that the trajectories of dust particles 
can be extended across the spacetime singularity that 
they cause to form. 

\vskip0.3cm
\noindent
(iii) Conservation of the total rest mass in the physical domain

\vskip0.3cm
From Eq.~(\ref{eq:D-evol}), we find that the rest mass 
of collapsing dust per unit Killing length $\sigma$ along 
$z$-coordinate in the extended domain, $-\infty<r<\infty$, 
is constant, where $\sigma$ is given by
\begin{equation}
\sigma=\int_{-\infty}^\infty D(t,r)dr. 
\end{equation}
By the condition (\ref{eq:PT-condition}), this can be rewritten in the form
\begin{equation}
\sigma=\int_0^\infty[D(t,r)+E(t,r)]dr. \label{eq:sigma-def}
\end{equation}
The above equation implies the conservation of the total rest mass in the 
physical domain, $r\geq0$. 

\begin{figure}[htbp]
\centering
\includegraphics[scale=0.5,angle=0]{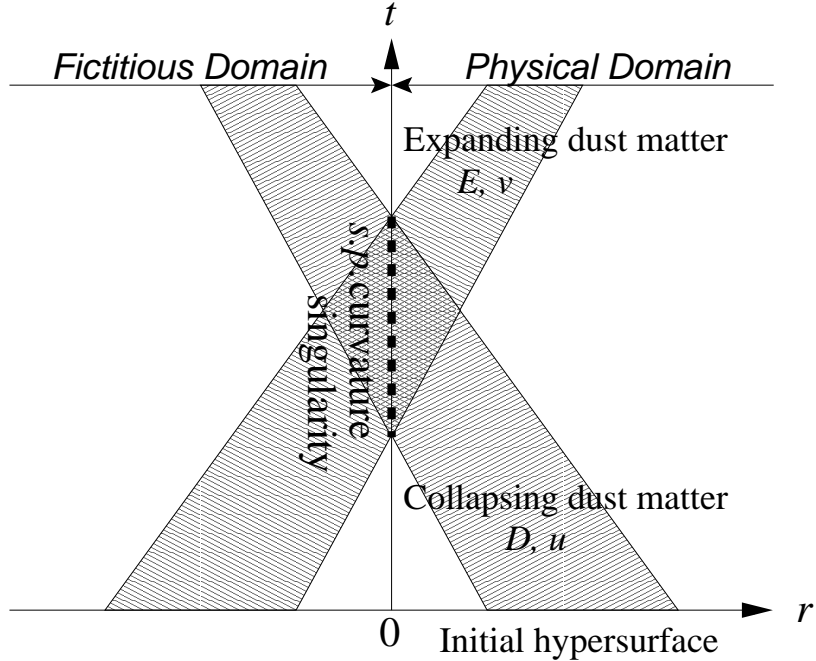}
\caption{A schematic diagram of the gravitational collapse of a thick 
cylindrical shell of dust. The collapsing dust forms a naked {\it s.p.} singularity, and then  dust particles pass through the naked singularity 
created by the dust itself. The regularity at the symmetry axis, $r=0$, 
is recovered after the dust leaves the naked singularity. 
}
\end{figure}

\vskip0.3cm
As demonstrated by our numerical results presented in the next section, 
if we set the initial conditions as $D\geq 0$ and $u\leq0$ 
with $E=0=v$ in the physical domain, 
all of the dust matter will enter into the fictitious domain, $r<0$, 
from the physical domain, $r\geq0$ (see Fig.~1). By the condition 
(\ref{eq:PT-condition}), the collapsing dust $(D,u)$ in the fictitious 
domain is equivalent to the expanding  
dust $(E,v)$ in the physical domain. Further, it should be noted that the 
completeness of radial geodesics implies that the dust matter can be regarded 
as a cold gas composed of collisionless particles; the collapsing dust $(D,u)$ 
passes through the spacetime singularity that it creates and then leaves 
the symmetry axis $r=0$ as expanding dust $(E,v)$. In this sense, 
it is natural to call the condition (\ref{eq:PT-condition}) 
{\it the passing-through condition}. As mentioned above, 
when the dust matter passes through the symmetry axis, the 
rest mass densities of dust, $\rho$ and $\mu$ and, thus, 
the Ricci scalar diverge there. 

\section{Numerical simulation}

\subsection{ Initial conditions}

We choose the initial conserved rest mass densities and 3-velocity 
fields to be  
\begin{eqnarray}
D&=&\frac{15\sigma}{32\pi w^5 l^5}
\left[r-l(1-w)\right]^2\left[r-l(1+w)\right]^2, \\
E&=&0, \\
u&=&-\sqrt{1-\exp\left(-\frac{\nu}{r}\right)}, \\
v&=&0,
\end{eqnarray}
for $l(1-w)<r<l(1+w)$ and vanishing elsewhere, where 
$\sigma$ is defined by Eq.~(\ref{eq:sigma-def}), 
$l$ and $w~(<1)$ are positive parameters controling the location 
and thickness of the dust shell, respectively, and $\nu$ is a parameter 
controling the gradient of the 3-velocity field, $u$. 

We determine the initial data of the metric variables $\beta$, 
$\gamma$ and $\psi$ 
and their time derivatives in the following manner. 
First, we set $\beta=1$ and $\dot{\beta}=0$. Then the constraint 
equations (\ref{eq:constraint-1}) and (\ref{eq:constraint-2}) become
\begin{eqnarray}
\gamma'&=&r{\psi'}^2+\frac{8\pi Ge^{\gamma-\psi}D}{\sqrt{1-u^2}}, 
\label{eq:dgamma}\\
\dot{\gamma}&=&-\frac{8\pi Ge^{\gamma-\psi}Du}{\sqrt{1-u^2}}.
\label{eq:dtgamma}
\end{eqnarray}
We have to integrate Eq.~(\ref{eq:dgamma}) numerically 
in order to obtain $\gamma$, while Eq.~(\ref{eq:dtgamma}) directly 
gives its time derivative. 
We are interested in an initial situation that is as close 
to the static configuration as possible with a non-vanishing 
initial collapsing velocity. Therefore we set $\dot{\psi}=0$. 
In order to determine the initial data of $\psi$, we use 
Eq.~(\ref{eq:psi-evol}) with $\ddot{\psi}=0$, i.e., 
\begin{equation}
\psi''=-\frac{1}{r}\left(\psi'+4\pi Ge^{\gamma-\psi}D\sqrt{1-u^2}\right).
\label{eq:dpsi}
\end{equation}
We numerically integrated Eqs.~(\ref{eq:dgamma}) and (\ref{eq:dpsi}) 
simultaneously outward from $r=0$ by imposing 
the boundary conditions $\gamma|_{r=0}=0$ and $\psi|_{r=0}=0=\psi'|_{r=0}$, 
since we assume that there is no spacetime singularity in the initial data. 

The vacuum region in the initial data obtained with the procedure 
described above agrees with the Levi-Civita solution, 
which is the unique solution for a vacuum static spacetime with 
whole cylinder symmetry,
\begin{equation}
\beta=1,
 ~~~~\gamma=\kappa^2 \ln r+\lambda~~~~{\rm and}~~~~\psi=-\kappa \ln r, 
\label{eq:LC}
\end{equation}
where $\kappa$ and $\lambda$ are parameters that characterize this solution. 
Integrating Eq.~(\ref{eq:dpsi}), we find that $\kappa$ vanishes for 
$r\leq l(1-w)$, whereas for $r\geq l(1+w)$, we have 
\begin{equation}
 \kappa = 4\pi G\int_{l(1-w)}^{l(1+w)}dr e^{\gamma-\psi}D\sqrt{1-u^2}. 
\end{equation}
Because $D$ is non-negative, $\kappa$ is positive in the domain $r\geq l(1+w)$.

\subsection{ Evolution}

\begin{figure}[htbp]
\centering
\includegraphics[scale=0.8,angle=0]{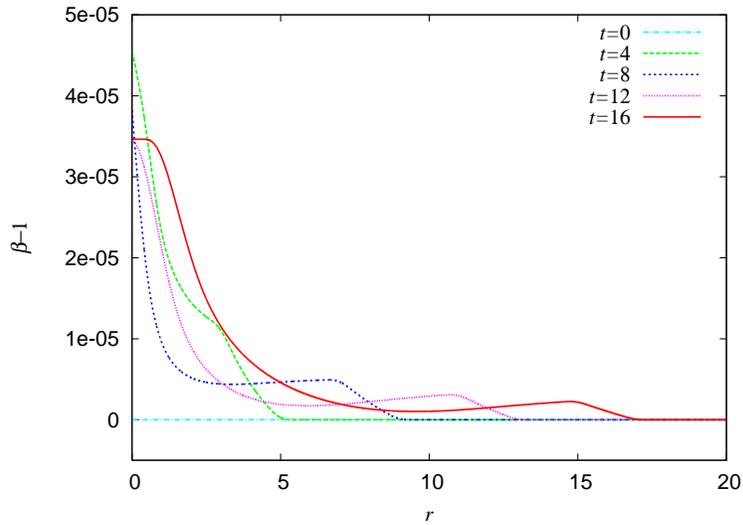}
\label{fig:beta}
\caption{Several snapshots of $\beta$ in the physical domain, $r\geq0$. 
The horizontal axis represents the radial coordinate, $r$.  
}
\end{figure}

\begin{figure}[htbp]
\centering
\includegraphics[scale=0.8,angle=0]{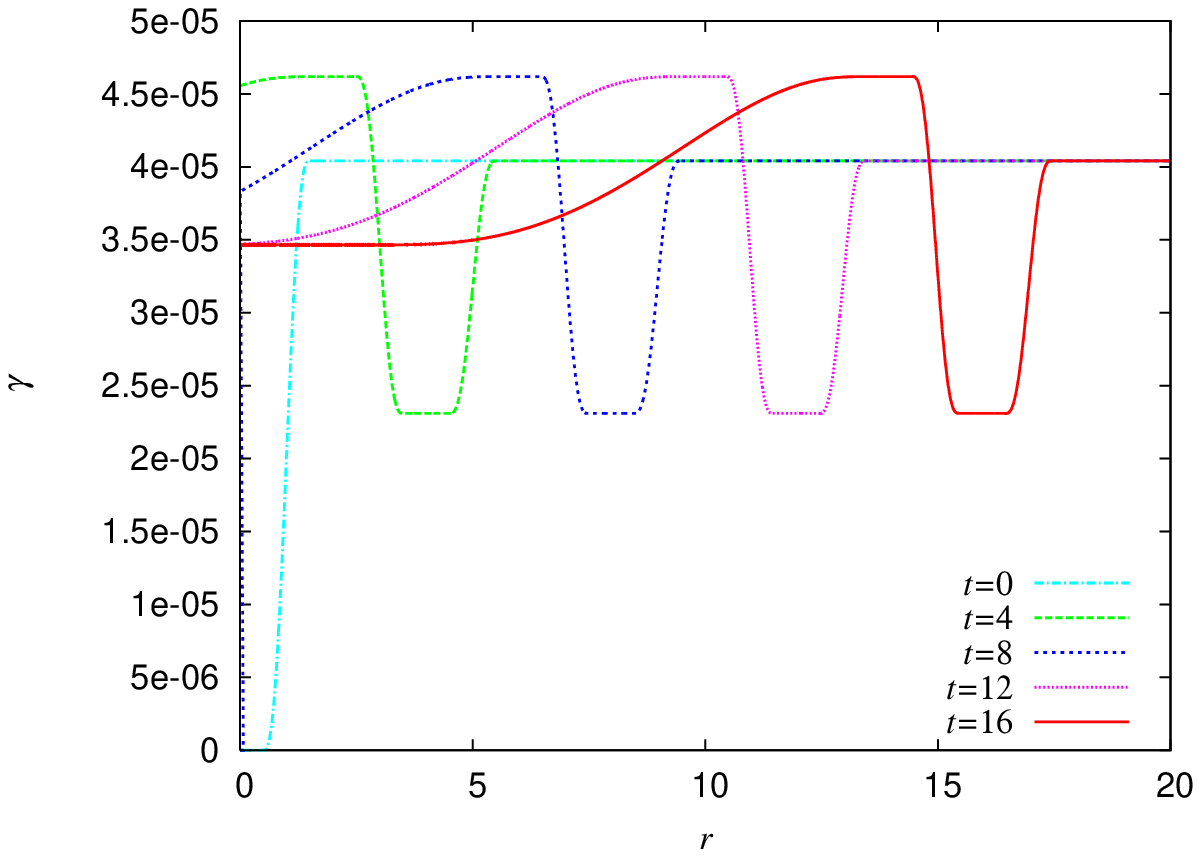}
\label{fig:gamma}
\caption{
The same as Fig.~2, but for $\gamma$. 
}
\end{figure}

\begin{figure}[htbp]
\centering
\includegraphics[scale=0.8,angle=0]{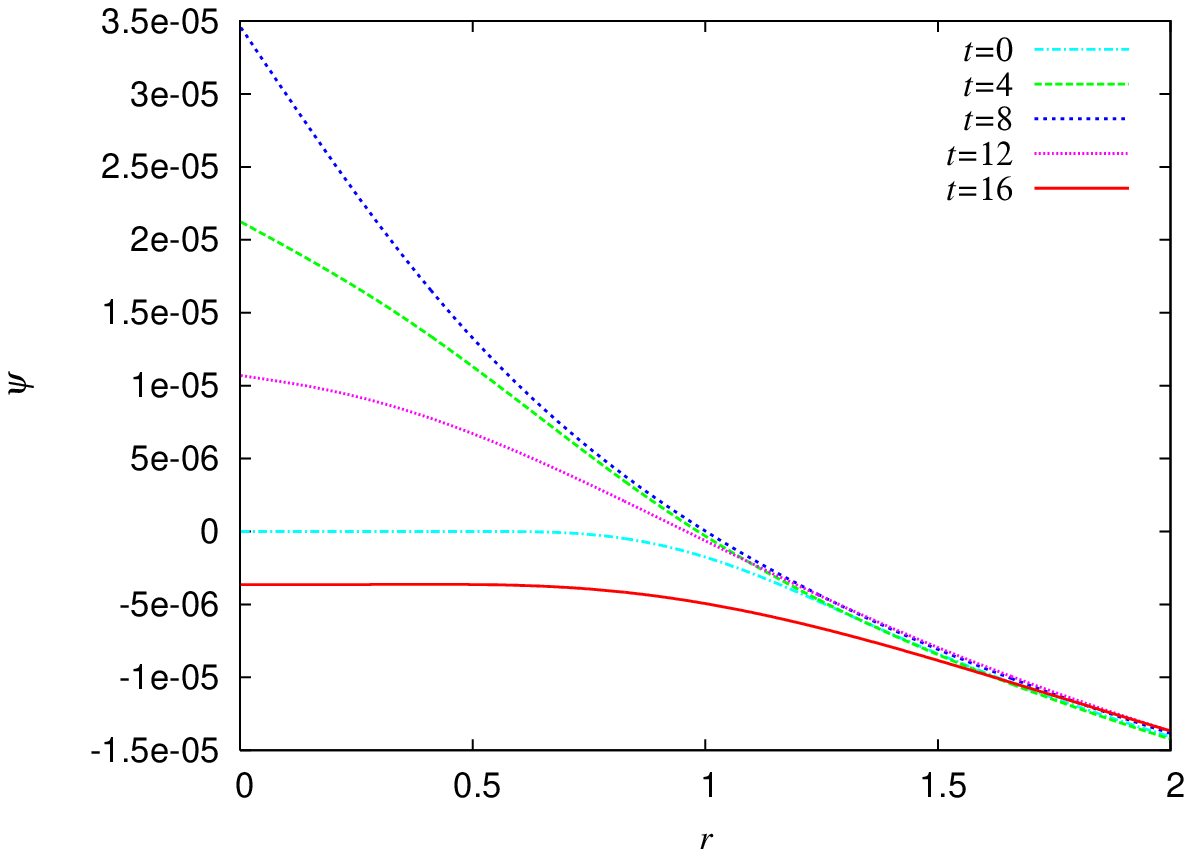}
\label{fig:psi}
\caption{
The same as Fig.~2, but for $\gamma$. 
}
\end{figure}

\begin{figure}[htbp]
\centering
\includegraphics[scale=0.8,angle=0]{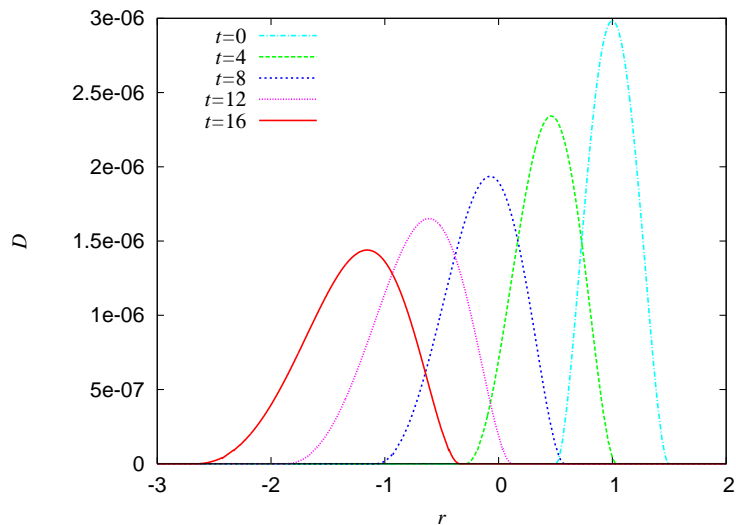}
\label{fig:density}
\caption{Several snapshots of the conserved rest mass density $D$ in the 
extended domain, $-\infty<r<\infty$. 
The horizontal axis represents the radial coordinate, $r$. 
}
\end{figure}

\begin{figure}[htbp]
\centering
\includegraphics[scale=0.8,angle=0]{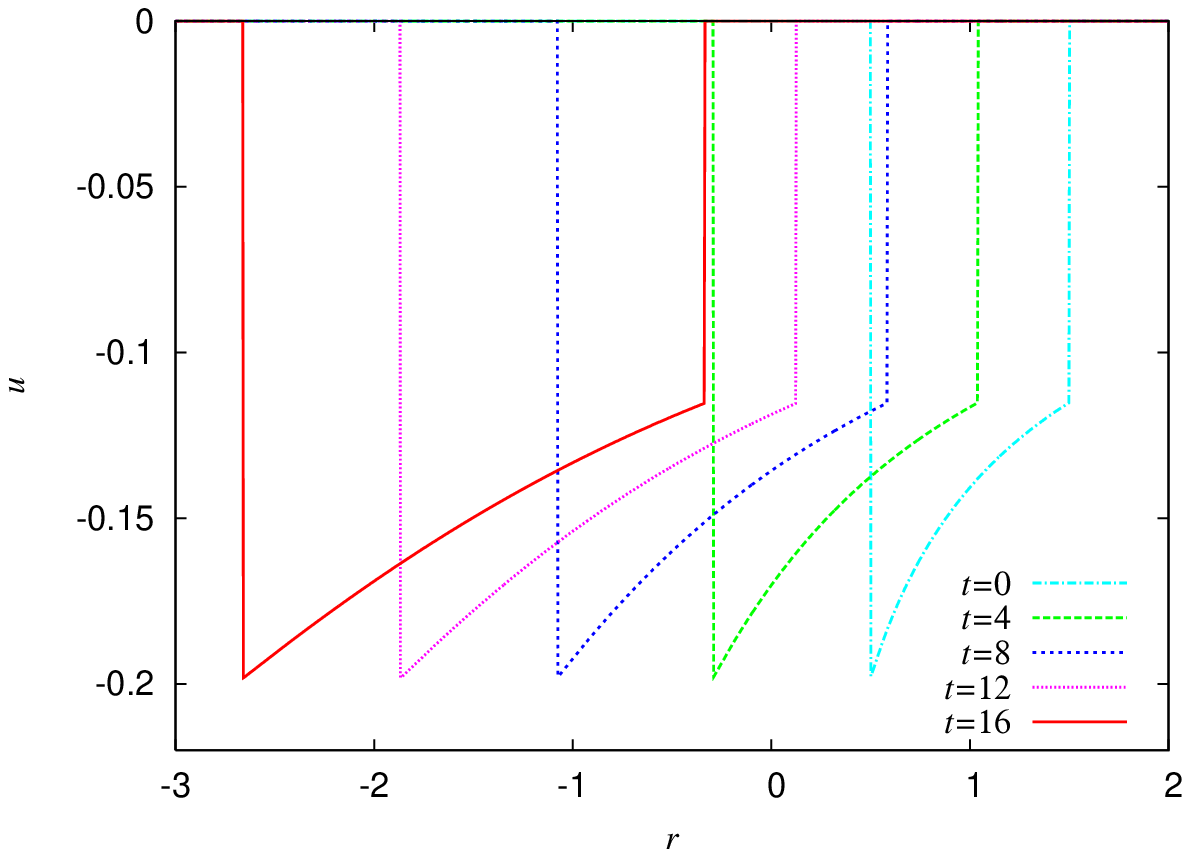}
\label{fig:velocity}
\caption{The same as Fig.~5, but for the velocity field  $u$. 
}
\end{figure}

In order to solve the evolution equations for the metric variables  
(\ref{eq:gamma-evol})--(\ref{eq:psi-evol}) numerically,  
we replaced the spatial derivatives by finite differences 
and adopted the MacCormack scheme for the time integration. 
In order to solve the evolution equations for the dust matter, we 
adopted the method proposed by Shapiro and Teukolsky,\cite{Shapiro:1991} 
with which we follow the motion of constituent mass 
shells of the dust matter moving along timelike geodesics. For the integration 
of the geodesic equations, we also adopted the MacCormack scheme. Then we  
constructed the conserved rest mass densities $D$ and $E$ and 
the velocity fields $u$ and $v$ from the positions and 
velocities of these mass shells. We assumed that 
all of the mass shells have the same rest mass per unit Killing length 
along $z$-coordinate. 
A linear combination of the constraint equations (\ref{eq:constraint-1}) and 
(\ref{eq:constraint-2}) was used to check the accuracy. 

We display examples of numerical solutions for the metric variables 
$\beta-1$, $\gamma$ and $\psi$ in the physical domain 
and the matter variables $D$ and $u$ in the extended domain in Figs.~2--6, 
respectively, where we have chosen the parameter values 
$l=1$, $w=0.5$, $\sigma=10^{-5}$ and $\nu=10^{-2}$. 
The numerically covered domain is $-20\leq r \leq20$. 
The numbers of spatial grid points and mass shells are 
$5\times10^3$ and $250$, respectively. 
In this numerical simulation, the spatially averaged relative 
error estimated from the constraint equations was always less than $10^{-8}$. 
Another criterion for the numerical accuracy is the difference of 
$\beta e^{-\gamma}|_{r=0}$ from unity. This difference was also always of 
order $10^{-8}$ in this simulation. Because $|\beta-1|$ and $\gamma$ are 
at most of order $10^{-5}$, the order of the relative error is 
understood to be $10^{-3}$. 

The boundary condition at the outer numerical boundary, $r=20$, is that 
none of the variables change there. This simplistic boundary 
condition is valid because we are interested in the evolution of 
the central region, and we stop the numerical simulations 
before the influence of the non-trivial initial configurations of the metric 
and matter variables reach the numerical boundary, $r=20$.  

It is seen from Figs.~2--4 that the metric variables $\beta$, 
$\gamma$ and $\psi$ are continuous and smooth in the numerically 
covered region. Further, it is seen 
from Figs.~5 and 6 that the conserved rest 
mass density $D$ is everywhere finite 
and that the absolute values of the velocity field $u$  
is always smaller than unity. However, as mentioned above, 
when the dust shell reaches the symmetry axis, 
$r=0$, the rest mass density $\rho$ diverges there, and thus 
a {\it s.p.} curvature singularity \cite{Hawking-Ellis} 
forms on the symmetry axis, $r=0$. The dust matter reaches the symmetry axis 
at $t\simeq3.0$ and forms a spacetime singularity. Then, all of the 
dust matter leaves the spacetime singularity, 
and regularity at the symmetry axis is recovered at $t\simeq12.3$ 
(see Fig.~5). The propagations of $\beta$ and $\gamma$ represent the 
gravitational radiation generated by the gravitational collapse 
of the dust matter. 

Because even after the appearance of the spacetime singularity, 
$\beta$, $\gamma$ and $\psi$ are everywhere 
finite and continuous, there is a causal future of this spacetime 
singularity, and there exists a Cauchy horizon associated with it. 
We can thus conclude that a naked singularity has indeed formed in the 
spacetime constructed by this numerical simulation. 

\section{ Completeness of causal geodesics}

In the spacetime of interest, 
since the metric tensor is $C^{1-}$ even in the neighborhood of 
the spacetime singularity, the parallel relation between 
any pair of vectors can be defined everywhere. 
A pair of spacelike vectors or timelike vectors $p^\mu$ and $q^\mu$ 
are parallel if and only if the equality  
\begin{equation}
\frac{g_{\mu\nu}p^\mu q^\nu}
{\sqrt{g_{\alpha\beta}p^\alpha p^\beta g_{\rho\sigma}q^\rho q^\sigma}}=\pm1
\end{equation}
is satisfied, where we have $+1$ for the pair of the spacelike vectors 
and $-1$ for the timelike vectors. 
In the case of a pair of null vectors $k^\mu$ and $l^\mu$, 
they are parallel if and only if  $g_{\mu\nu}k^\mu l^\nu=0$ and 
the product of their time components are positive. 
Further, in the spacetime of interest, the Christoffel symbols 
in the Cartesian coordinate system defined by Eq.~(\ref{eq:Cartesian}) 
have limits with directional dependence for $r\rightarrow0$, but these 
limits are finite (see Appendix C). 
By virtue of these two properties of the spacetime singularity at $r=0$, 
almost all causal geodesics are complete, 
even if they intersect this spacetime singularity. 
We show how to construct complete solutions for the 
geodesic equations below. 

In the coordinate system (\ref{eq:cylindrical}), 
the geodesic equations take the forms 
\begin{eqnarray}
\frac{d}{d\tau}\left(e^{2(\gamma-\psi)}\frac{dt}{d\tau}\right)
&=&\left(\dot{\gamma}-\dot{\psi}\right)e^{2(\gamma-\psi)}
\left[
\left(\frac{dt}{d\tau}\right)^2
-\left(\frac{dr}{d\tau}\right)^2
\right]
\nonumber \\
&-&\left(\frac{\dot{\beta}}{\beta}-\dot{\psi}\right)
r^2\beta^2e^{-2\psi}
\left(\frac{d\varphi}{d\tau}\right)^2
-\dot{\psi}e^{2\psi}\left(\frac{dz}{d\tau}\right)^2, \label{eq:t-geodesic} \\
&& \nonumber \\
\frac{d}{d\tau}\left(e^{2(\gamma-\psi)}\frac{dr}{d\tau}\right)
&=&-\left(\gamma'-\psi'\right)e^{2(\gamma-\psi)}
\left[
\left(\frac{dt}{d\tau}\right)^2
-\left(\frac{dr}{d\tau}\right)^2
\right]
\nonumber \\
&+&\left(\frac{1}{r}+\frac{\beta'}{\beta}-\psi'\right)
r^2\beta^2e^{-2\psi}
\left(\frac{d\varphi}{d\tau}\right)^2
+\psi'e^{2\psi}\left(\frac{dz}{d\tau}\right)^2, 
\label{eq:r-geodesic} \\
&& \nonumber \\
\frac{d}{d\tau}\left(r^2\beta^2e^{-2\psi}\frac{d\varphi}{d\tau}\right)
&=&0, \label{eq:phi-geodesic} \\
&& \nonumber \\
\frac{d}{d\tau}\left(e^{2\psi}\frac{dz}{d\tau}\right)
&=&0, \label{eq:z-geodesic}
\end{eqnarray}
where $\tau$ is the affine parameter. The third and fourth equations 
can be easily integrated once, and we have
\begin{equation}
r^2\beta^2e^{-2\psi}\frac{d\varphi}{d\tau}=L,~~~~~~
e^{2\psi}\frac{dz}{d\tau}=P, \label{eq:c-quantities}
\end{equation}
where $L$ and $P$ are constants of integration. 
There are three categories of causal geodesics, which 
behave in very different manners near 
the spacetime singularity. The first category consists of 
the causal geodesics with $r=0=dr/d\tau$. 
We call these geodesics {\it central geodesics}. 
The second category consists of the causal geodesics with vanishing 
$L$ but non-vanishing $dr/d\tau$. We call causal 
geodesics in this second category {\it radial geodesics}. The third 
category consists of geodesics which have non-vanishing $L$. 
We call these {\it non-radial geodesics}. 
As shown in Appendix B, non-radial geodesics cannot reach the symmetry 
axis, $r=0$. For this reason, hereafter we focus on the central and radial geodesics. 

\subsection{Radial geodesics}

First, we consider radial geodesics. Hereafter we assume that all 
causal geodesics are future directed, i.e., 
their time components are positive. 
An extension of a radial geodesic beyond the spacetime singularity at $r=0$ 
is also a radial geodesic. Thus we consider two radial geodesics, 
denoted by $\bm{k}_{-}$ and $\bm{k}_{+}$. 
We assume that $\bm{k}_{-}$ intersects the spacetime singularity at $t=t_0$ 
and that $\bm{k}_{+}$ leaves the spacetime singularity at $t=t_0$ 
Thus $\bm{k}_{+}\cup\bm{k}_{-}$ consists of a continuous curve.  
In order for $\bm{k}_{+}$ to be a unique extension of $\bm{k}_{-}$ beyond the 
spacetime singularity, the tangent vectors of these two geodesics should be 
identical at the spacetime singularity, since the geodesic 
tangent vector is parallelly transported along the geodesic, by definition. 
If this condition uniquely determines $\bm{k}_{+}$ for any $\bm{k}_{-}$, 
we can conclude that all of the radial geodesics are complete in this spacetime. 

In general, a radial geodesic is specified by $\varphi=$[constant]. 
We denote the angle coordinates of $\bm{k}_{\pm}$ by $\varphi_{\pm}$.
The components of their tangent vectors, $k^\mu_{\pm}$, 
in the cylindrical coordinate system are written in the form
\begin{equation}
k_{\pm}^\mu=\left(\sqrt{-e^{2(\psi-\gamma)}\chi+V_{\pm}^2+e^{-2\gamma}P_{\pm}^2}
,~V_{\pm},~0,~e^{-2\psi}P_{\pm}\right)_{tr\varphi z},
\end{equation}
where $V_{\pm}$ are functions of the 
affine parameter $\tau$, $P_{\pm}$ is constant, 
and $\chi$ is $-1$ if $\bm{k}_{\pm}$ are timelike geodesics, 
while $\chi$ vanishes 
if $\bm{k}_{\pm}$ are null. Here we adjust the affine parameter $\tau$ so that 
the geodesic $\bm{k}_{-}$ intersects the spacetime singularity at $\tau=0$, 
while $\bm{k}_{+}$ leaves the spacetime singularity at $\tau=0$, or, 
in other words, a negative $\tau$ is assigned to $\bm{k}_{-}$ 
and a positive $\tau$ is assigned to $\bm{k}_{+}$. By assumption, we have 
\begin{equation}
\lim_{\tau\rightarrow0}V_{-}<0~~~~{\rm and}~~~~\lim_{\tau\rightarrow0}V_{+}>0.
\label{eq:V-limit}
\end{equation}
In order to see the continuity of the tangent vectors at  
the spacetime singularity, the Cartesian coordinate system defined by 
Eq.~(\ref{eq:Cartesian}) is useful, because the Cartesian coordinate 
system covers the symmetry axis, $r=0$, 
whereas the cylindrical coordinate system 
does not, and the components of vectors in this coordinate system 
cannot be specified at $r=0$. 
The components of the tangent vectors of $\bm{k}_{\pm}$ 
in the Cartesian coordinate system are therefore given by 
\begin{equation}
k_{\pm}^\mu=\left(\sqrt{-e^{2(\psi-\gamma)}\chi+V_{\pm}^2+e^{-2\gamma}P_{\pm}^2},
~V_{\pm}\cos\varphi_{\pm},~V_{\pm}\sin\varphi_{\pm},~e^{-2\psi}P_{\pm}\right)_{txyz}. 
\end{equation}
Since the metric variables $\beta$, $\gamma$ and 
$\psi$ are $C^{2-}$ functions of $t$ and $r$, the quantities $V_{\pm}$  
have finite limits as $\tau\rightarrow 0$. 
The components of the tangent vectors $k^\mu_{\pm}$ approach identical values  
in the limit that the spacetime singularity is approached, 
$\tau\rightarrow0$, 
if and only if the following conditions are satisfied of these: 
\begin{eqnarray}
P_{+}=P_{-},~~
\lim_{\tau\rightarrow0}V_{+}=-\lim_{\tau\rightarrow0}V_{-}~~~~{\rm and}~~~~
\varphi_{+}=\varphi_{-}+\pi.
\end{eqnarray}
The first of these equations is obtained from the continuity of the $z$-component, 
whereas the second one is obtained from the continuity of the $t$-component and 
Eq.~(\ref{eq:V-limit}). The final one is derived from the continuities of 
the $x$- and $y$-components. 
Here it should be noted that because the metric tensor is defined 
even at the spacetime singularity at $r=0$, the fact that the 
limiting values of the components $k_{\pm}^\mu$ obtained as 
$\tau\rightarrow0$ are identical implies that the tangent 
vectors themselves coincide at the spacetime singularity. 
Since the first order derivatives of $\gamma$ and $\psi$ are $C^{1-}$ functions 
of $t$ and $r$ in the domain $0\leq r<\infty$, the standard existence theorems 
for ordinary differential equations can be applied to the geodesic equations 
for radial geodesics $\bm{k}_{+}$, 
\begin{eqnarray}
\frac{dV_{+}}{d\tau}
&=&
e^{2(\psi-\gamma)}\left[(2\psi'-\gamma')e^{-2\psi}P_{+}^2
+(\dot{\psi}-\dot{\gamma})\chi\right]+2V_{+}\frac{d}{d\tau}(\psi-\gamma), \\
\frac{dt}{d\tau}&=&\sqrt{-e^{2(\psi-\gamma)}\chi+V_{+}^2+e^{-2\gamma}P_{+}^2}, \\
\frac{dr}{d\tau}&=&V_{+}, 
\end{eqnarray}
and Eq.~(\ref{eq:c-quantities}) with $L=0$ and $P=P_{+}$, 
in order to show that the geodesic $\bm{k}_{+}$ 
is uniquely determined by $P_{+}$, $V_{+}$ and $\varphi_{+}$ at $(t,r)=(t_0,0)$. 
Therefore we conclude that $\bm{k}_{+}\cup\bm{k}_{-}$ is the unique extension of 
the geodesic $\bm{k}_{-}$ beyond the spacetime singularity. 

\subsection{Central geodesics}

In order to elucidate the extendibility of the central geodesics, first we consider 
a causal curve along $r=0$. By definition, the tangent vector to this 
curve is given by  
\begin{equation}
\frac{dx^\mu}{d\lambda}=\left(\frac{dt}{d\lambda},~0,~0,~\frac{dz}{d\lambda}\right).
\end{equation}
Thus we have
\begin{eqnarray}
\Gamma^x_{\mu\nu}\frac{dx^\mu}{d\lambda}\frac{dx^\nu}{d\lambda}&=&
\left[(\gamma'-\psi')\left(\frac{dt}{d\lambda}\right)^2
-\psi'e^{-2\gamma+4\psi}\left(\frac{dz}{d\lambda}\right)^2\right]\cos\varphi, \\
\Gamma^y_{\mu\nu}\frac{dx^\mu}{d\tau}\frac{dx^\nu}{d\lambda}&=&
\left[(\gamma'-\psi')\left(\frac{dt}{d\lambda}\right)^2
-\psi'e^{-2\gamma+4\psi}\left(\frac{dz}{d\lambda}\right)^2\right]\sin\varphi, 
\end{eqnarray}
where again we note that $\varphi$ is arbitrary at $r=0$. 
If the symmetry axis is regular, $\gamma'$ and $\psi'$ vanish there, and thus 
the above quantities vanish. In this case, this curve can be a geodesic, since 
$d^2x/d\lambda^2$ and $d^2y/d\lambda^2$ vanish, as found from the geodesic equations. 
By contrast, if a spacetime singularity appears on the symmetry axis, $\gamma'$ 
and $\psi'$ do not vanish there, and therefore the above quantities become  
indefinite, due to the arbitrariness of $\varphi$. 
This implies that the curve $x=y=0=dx/d\lambda=dy/d\lambda$ 
cannot be a geodesic at the spacetime singularity, 
since the geodesic equations are not well-defined there. 
However, it should be noted that the geodesics along $r=0$ cannot be  
extended across the spacetime singularity as a causal {\it geodesic}, 
but they can be extended as a causal {\it curve}, since the metric tensor 
is defined even at the spacetime singularity.\footnote{The same is true for the 
null dust  case. Two of the present authors have argued that a spacetime 
exhibiting the gravitational collapse of cylindrical null dust 
is geodesically complete 
if the passing-through condition is imposed,\cite{Kurita_N:2006} 
but this argument is not exact. } 
In this sense, this singularity is 
very weak, like singular hypersurfaces, as discussed below. 

\subsection{ Comparison of the present spacetime singularity with others}

Here it is worthwhile considering the case that $r=0$ is conically singular, i.e., 
$e^{2\gamma}\neq\beta^2$ at $r=0$. Also in this case, we can uniquely determine 
a geodesic $\bm{k}_{+}$ for a given causal geodesic $\bm{k}_{-}$ in the same 
manner as in $\S$ 5.2. 
However, since, in contrast to the cases of interest presently, 
the metric tensor is not defined at the conical 
singularity, $r=0$, we cannot conclude the vectors are parallel there. 
Hence, we cannot claim that the tangent vectors to $\bm{k}_-$ and 
$\bm{k}_+$ are parallel to each other at the conical singularity at $r=0$, 
even if their components with respect 
to the Cartesian coordinate system are identical to each other in the limit 
$\tau\rightarrow0$.  Parallel transport across the conical singularity 
is impossible, and therefore the geodesics that intersect the conical singularity 
at $r=0$ are not extendible as continuous geodesics. 

As shown in $\S\S$ 5.1 and 5.2, 
although all of the radial geodesics are complete, 
the central geodesics are not complete in the spacetime of interest. 
Thus, this spacetime is geodesically 
incomplete. However, this spacetime singularity is not too 
strong to treat within the framework of general relativity. 
The same situations can also be realized in the case of singular hypersurfaces, 
which can be treated within the framework of general relativity with 
a clear physical meaning.\cite{Israel}
To make the discussion clear, let us consider a static spherically 
symmetric shell with infinitesimal thickness but finite mass, 
which constitutes a singular timelike hypersurface. From a physical point of view, 
such a thin shell is supported by the tangential pressure against the 
gravitational force action to collapse it. 
We assume that it exists in a vacuum spacetime. 
By the assumed symmetry, the inside of the singular hypersurface is 
described by the Minkowski geometry, 
\begin{equation}
ds^2=-dt^2+dr^2+r^2(d\theta^2+\sin^2\theta d\varphi^2),
\end{equation}
whereas the outside is described by the Schwarzschild geometry,
\begin{equation}
ds^2=-\left(1-\frac{r_{\rm g}}{r}\right)dT^2
+\left(1-\frac{r_{\rm g}}{r}\right)^{-1}dr^2+r^2(d\theta^2+\sin^2\theta d\varphi^2),
\end{equation}
where $r_{\rm g}$ is a non-vanishing constant smaller than the areal radius of 
the shell, $r=r_{\rm shell}$. 
Then, consider curves with constant spatial coordinates  
$r$, $\theta$ and $\varphi$. They are timelike geodesics in the Minkowski domain, 
but they are not in the Schwarzschild domain. 
However, we cannot determine whether or not a timelike 
curve attached to this spherical shell, 
$r=r_{\rm shell}$, is a timelike geodesic, because the 
connection is not defined on the spherical shell. 
This situation is similar to that of the central geodesics in 
the present case. 
 
\section{ Curvature strength}

In order to investigate the strength of the curvature at the spacetime 
singularity, $r=0$, we consider timelike geodesics that intersect the 
spacetime singularity. The curvature 
strength of spacetime singularities was defined in the hope that the 
weak convergence of 
geodesic congruences would reveal the extendibility of the spacetime in 
a distributional sense.\cite{Clarke} In this context, Tipler 
defined the {\it strong curvature condition},\cite{Tipler:1977} 
and Kr\'{o}lak defined a weaker condition 
called the {\it limiting focusing condition}.\cite{Krolak:1987}

As mentioned above, non-radial geodesics cannot intersect the spacetime  
singularity, and therefore we consider only 
radial and central ones. The components of the radial or central geodesic tangent 
$e_{(0)}^\mu$ are given by 
\begin{equation}
e_{(0)}^\mu=\left(\frac{dt}{d\tau},\frac{dr}{d\tau},0,e^{-2\psi}P\right).
\end{equation}
Then we have an orthonormal frame parallely propagating along this timelike geodesic, 
\begin{eqnarray}
e^\mu_{(1)}&=&\frac{1}{N}\left(\frac{dr}{d\tau},\frac{dt}{d\tau},0,0\right), \\
e^\mu_{(2)}&=&\left(0,0,\frac{e^{2\psi}}{r\beta},0\right), \\
e^\mu_{(3)}&=&\frac{e^{-\psi}}{N}\left(P\frac{dt}{d\tau},P\frac{dr}{d\tau},0,
N^2\right),
\end{eqnarray}
where 
\begin{equation}
N:=\sqrt{1+e^{-2\psi}P^2}.
\end{equation}
The strength of the tidal force measured by an observer moving along
this timelike geodesic is given by the components 
of the Riemann tensor with respect to the above tetrad basis, 
$R_{(0)(A)(0)(B)}$ ($A,~B=1,2,3$). 

We give the coordinate basis components of the Riemann tensor in Appendix D. 
Using those components, it can be seen that the only $R_{(0)(2)(0)(2)}$ 
becomes infinite at the spacetime singularity, $r=0$, and it is given by 
\begin{eqnarray}
R_{(0)(2)(0)(2)}&=&\frac{1}{r}\Biggl[(\gamma'-\psi')
\left(\frac{dt}{d\tau}\right)^2
+2\left(\dot{\gamma}-\frac{\dot{\beta}}{\beta}\right)\frac{dt}{d\tau}\frac{dr}{d\tau}
\nonumber \\
&+&\left(\gamma'+\psi'-\frac{2\beta'}{\beta}\right)\left(\frac{dr}{d\tau}\right)^2
-e^{-2\gamma}P^2\psi'
\Biggr]
\nonumber \\
&+&\left[
\ddot{\psi}-\dot{\gamma}\dot{\psi}
-\gamma'\psi'+{\psi'}^2+\frac{\dot{\beta}}{\beta}(\dot{\gamma}+\dot{\psi})
+\frac{\beta'}{\beta}\left(\gamma'-\psi'\right)
-\frac{\ddot{\beta}}{\beta}\right]
\left(\frac{dt}{d\tau}\right)^2
\nonumber \\
&+&2
\left[
\dot{\psi}'-\gamma'\dot{\psi}
-\dot{\gamma}\psi'+\psi'\dot{\psi}+\frac{\dot{\beta}}{\beta}\gamma'
+\frac{\beta'}{\beta}\dot{\gamma}
-\frac{\dot{\beta}'}{\beta}
\right]
\frac{dt}{d\tau}\frac{dr}{d\tau} 
\nonumber \\
&+&\left[
\psi''-\dot{\gamma}\dot{\psi}
-\gamma'\psi'+\dot{\psi}^2+\frac{\dot{\beta}}{\beta}(\dot{\gamma}-\dot{\psi})
+\frac{\beta'}{\beta}\left(\gamma'+\psi'\right)
-\frac{\beta''}{\beta}\right]\left(\frac{dr}{d\tau}\right)^2
\nonumber \\
&-&e^{-2\gamma}P^2
\left(
\dot{\psi}^2
-{\psi'}^2
-\frac{\dot{\beta}}{\beta}\dot{\psi}
+\frac{\beta'}{\beta}\psi'
\right). \label{eq:Riemann}
\end{eqnarray}
As in the previous section, the affine parameter $\tau$ is adjusted 
so that the geodesics intersect the singularity at $\tau=0$. 

Before the singularity formation, $\beta'$, $\gamma'$ and $\psi'$ vanish 
at the symmetry axis, $r=0$, and thus along the central geodesics, $R_{(0)(2)(0)(2)}$ 
is given by
\begin{eqnarray}
&&R_{(0)(2)(0)(2)}=e^{-2(\gamma-\psi)}
\left[
\ddot{\psi}-\dot{\gamma}\dot{\psi}
-\gamma'\psi'+{\psi'}^2+\frac{\dot{\beta}}{\beta}(\dot{\gamma}+\dot{\psi})
+\frac{\beta'}{\beta}\left(\gamma'-\psi'\right)
-\frac{\ddot{\beta}}{\beta}
\right]
\nonumber \\
&&-e^{-2\gamma}P^2
\left[
\ddot{\psi}-\dot{\gamma}\dot{\psi}
-\gamma'\psi'-\dot{\psi}^2+2{\psi'}^2
+\frac{\dot{\beta}}{\beta}(\dot{\gamma}+2\dot{\psi})
+\frac{\beta'}{\beta}\left(\gamma'-2\psi'\right)
-\frac{\ddot{\beta}}{\beta}
\right],
\end{eqnarray}
where we have used the normalization of the geodesic tangent. 
Because the metric functions $\beta$, $\gamma$ and $\psi$ are 
regular with respect to $t$ and $r$ on the symmetry axis, $r=0$, 
and because $\beta$ must be positive, 
we can easily see that $R_{(0)(2)(0)(2)}$ 
is finite in the limit $\tau\rightarrow0-$. It suddenly diverges 
just when the conserved densities $D$ and $E$ become positive at $r=0$. 
Thus, neither the strong curvature condition nor the  
limiting focusing condition is satisfied along the 
central geodesics.\cite{Clarke} 

We can see from Eq.~(\ref{eq:Riemann}) that 
in the case of a radial geodesic with $dr/d\tau<0$, 
the component $R_{(0)(2)(0)(2)}$ takes the form
\begin{equation}
R_{(0)(2)(0)(2)}=\frac{I(\tau)}{r(\tau)}+J(\tau),
\end{equation}
where $I$ and $J$ are $C^{1-}$ functions of the affine parameter $\tau$. 
Thus we have
\begin{equation}
Z:=\lim_{\tau\rightarrow0}\tau \bigl|R_{(0)(2)(0)(2)}\bigr| 
=\lim_{\tau\rightarrow0}
\biggl|I(\tau)
\left(\frac{dr}{d\tau}\right)^{-1}
\biggr|,
\end{equation}
where we have used l'Hopital's theorem in the final equality, 
\begin{equation}
\lim_{\tau\rightarrow0}\frac{\tau}{r}=\lim_{\tau\rightarrow}
\left(\frac{dr}{d\tau}\right)^{-1}. 
\end{equation}
Further, by l'Hopital's theorem, we have
\begin{eqnarray}
&&\lim_{\tau\rightarrow0}
\frac{\int_{c_1}^\tau d\bar{\tau}\int_{c_2}^{\bar\tau}d\hat{\tau}
\bigl|R_{(0)(2)(0)(2)}(\hat{\tau})\bigr|}{\tau\ln\tau-\tau}
=\lim_{\tau\rightarrow0}
\frac{\int_{c_2}^\tau d\hat{\tau}
\bigl|R_{(0)(2)(0)(2)}(\hat{\tau})\bigr|}{\ln\tau} =Z, \label{eq:limit}
\end{eqnarray}
where $c_1$ and $c_2$ are negative constants. 
Thus we find
\begin{equation}
\lim_{\tau\rightarrow0}\int_{c_1}^\tau d\bar{\tau} \int_{c_2}^{\bar{\tau}}d\hat{\tau}
\bigl|R_{(0)(2)(0)(2)}\bigr|=Z\lim_{\tau\rightarrow 0}
(\tau\ln\tau-\tau)=0.
\end{equation}
The above equation implies that this spacetime singularity is not the strong curvature 
singularity with respect to the timelike geodesics.\cite{Clarke} We can see that 
the same is true for null geodesics by using the same procedure as above. 

However, this spacetime singularity satisfies the limiting focusing 
condition, since from Eq.~(\ref{eq:limit}), we have
\begin{equation}
\int_{c_2}^\tau d\hat{\tau}
\bigl|R_{(0)(2)(0)(2)}(\hat{\tau})\bigr|
=Z\lim_{\tau\rightarrow0}\ln\tau=\infty.
\end{equation}
In this sense, this singularity is slightly stronger  
than singular hypersurfaces. 
This is an example of a spacetime that is extendible across the 
spacetime singularity which satisfies the limiting focusing condition.\cite{Clarke} 

\section{ Summary and discussion}

We have studied the gravitational collapse of a cylindrical thick shell,
in other words, a hollow cylinder, composed of  
dust matter and presented a physically reasonable 
boundary condition at the resultant string-like naked 
singularity formed at the symmetry axis. With this boundary condition, 
the trajectories of dust particles can be extended across the naked 
singularity, and thus the dust matter can be regarded as a cold gas of 
collisionless particles of infinitesimal mass; 
the collapsing dust matter passes through the 
naked singularity created by the dust 
itself without any interaction between individual 
constituent particles. When the dust matter leaves this region, 
the naked singularity disappears, and the regularity at the symmetry axis is 
recovered. We performed numerical simulations under this passing-through 
boundary condition and showed that the picture described above is accurate. 
The obtained spacetime should represent a phenomenological description of
such a physically realistic system of collisionless particles.

This naked singularity is a scalar polynomial curvature singularity. 
The strength of the curvature at the naked singularity was investigated, 
and we found that it does not satisfy the strong curvature condition defined by Tipler 
but that it does satisfy the limiting focusing condition defined by Kr\'{o}lak. 
A causal geodesic along the symmetry axis, $r=0$, which is called the central 
geodesic, cannot be extended across the spacetime singularity 
as a causal {\it geodesic}, but it can be extended as a causal {\it curve}. 
All the other causal geodesics that intersect the naked singularity 
can be extended across the naked singularity. Because the central geodesics 
have zero measure in the space of solutions for the geodesic equations, 
we can conclude that the extended spacetime is complete for 
almost all of the causal geodesics. The present results imply 
that gravity produced by a cylindrical thick shell of  
dust matter is too weak to bind the shell, 
even if it engenders a curvature singularity.
We conjecture that this weakness of gravity is the 
origin of its nakedness. Such a naked singularity does not lead to 
serious problems, because we can describe this system with a $C^{1-}$ 
spacetime, as in the case of singular hypersurfaces. 

In the case of a solid dust cylinder, the spacetime 
singularity formed by its gravitational 
collapse will be stronger than in the case of a hollow cylinder 
for the following reason. A regular solid dust cylinder should satisfy
the relations
\begin{equation}
\lim_{r\rightarrow0}\frac{D}{r}>0. ~~~~{\rm and}~~~~
\lim_{r\rightarrow0}\frac{u}{r}=u_{\rm c}, 
\end{equation}
where $u_{\rm c}$ is a real number. 
Because a collapsing solid dust cylinder will have negative $u_{\rm c}$, 
$u$ will be monotonically decreasing in a sufficiently 
small neighborhood of the symmetry axis, $r=0$. 
This implies that a caustic of dust matter appears at $r=0$ within a finite time, 
and $D$ diverges there. The resulting spacetime singularity will be of the 
third kind defined in $\S$ 3.2. In this case, provided that 
the boundary conditions 
(\ref{eq:beta-bc})--(\ref{eq:psi-bc}) are imposed, $\beta'$, $\gamma'$ and $\psi'$ 
diverge at $r=0$, and therefore the metric tensor cannot be $C^{1-}$
for any extended spacetime.
Thus, we need another prescription to deal with this spacetime 
singularity. Here, it is worth noting that if an infinitesimal 
cylindrical portion of dust matter is removed from around the center of 
this solid dust cylinder, it becomes a cylindrical thick shell, and hence 
the spacetime singularity becomes so weak that almost all causal geodesics 
are complete. This means that the strength of the spacetime 
singularity comes from the infinitesimal central portion of the dust 
cylinder, which seems to be insignificant from a physical point of view. 
Therefore, the spacetime singularity formed by a solid dust cylinder also seems to  
be so weak that we could find an appropriate prescription 
to treat this spacetime singularity. This issue will be discussed elsewhere. 

The Shapiro-Teukolsky singularity, which was reported to be a naked
singularity formed in the spindle 
collapse of collisionless particles, might be similar to that studied in 
the present case, except in the neighborhoods of the poles of 
the spindle matter distribution. In general, if
the naked singularity formed by the highly elongated gravitational 
collapse is so weak as to allow $C^{1-}$ extension of spacetime,
it can also be regarded as phenomenological, 
employing coarse-graining of `microphysics', such as 
infinitesimally massive collisionless particles.

In the present investigation, we did not continue the numerical simulations until 
nearly equilibrium configurations were realized. Due to the pressureless nature 
of dust matter, shell crossing singularities must appear. In the numerical 
scheme adopted in this paper, the dust matter is treated as a system composed 
of collisionless particles, and thus, in principle, we are able to follow the dynamics 
even after the appearance of shell crossing singularities in our numerical 
code, as long as they do not appear on the symmetry axis, $r=0$. 
This system might generate a large amount of 
gravitational radiation, as in the case of the adiabatic gravitational collapse of 
a cylindrically symmetric ideal gas. \cite{Piran:1978} 
Investigating this point is also a future work. 

\section*{Acknowledgements}

We are grateful to H.~Ishihara and our colleagues in the astrophysics and 
gravity group at Osaka City University for useful and helpful 
discussions and criticism. This work was partially supported by 
Grants-in-Aid for Scientific Research (C), No.16540264, and 
for Young Scientists (B), No.18740144 from the 
Ministry of Education, Science, Sports and Culture of Japan. 	

\appendix
\section{The Positivity of $\beta$} 

In order to demonstrate the relation $\beta>0$, we invoke the 
following theorem proven by Hayward \cite{Hayward:2000}: 
{\it If the dominant energy condition is satisfied, the $C$-energy 
is non-decreasing in outward achronal direction in untrapped regions.}
The dominant energy condition 
for the matter of the stress-energy tensor $T_{\mu\nu}$ 
is the condition that $T_{\mu\nu}V^\mu$ be a causal vector for an arbitrary causal 
vector $V^\mu$. 
Here, {\it untrapped} means that the relation $\dot{R}^2-{R'}^2<0$ holds, where 
\begin{equation}
R:=\sqrt{g_{\varphi\varphi}g_{zz}}=r\beta
\end{equation}
is the area element in the 2-space labeled by $t$ and $r$. 
The $C$-energy $C$ is a quasi-local energy per unit Killing length along 
$z$-coordinate included within the cylinder of radius $r$, \cite{Thorne:1965} 
\begin{equation}
C:=\frac{1}{8G}\left[1+e^{-2\gamma}(\dot{R}^2-{R'}^2)\right].
\end{equation}
By the above definition, a non-decreasing 
$C$-energy for the outward achronal direction implies the relation  
\begin{equation}
\xi^\mu\frac{\partial}{\partial x^\mu} 
\left[e^{-2\gamma}(\dot{R}^2-{R'}^2)\right] \geq0, \label{eq:hayward}
\end{equation}
where $\xi^\mu$ is an arbitrary outward acrhonal vector. 

In the present paper, we consider a spacetime that is identical to 
the Levi-Civita spacetime in the spacelike asymptotic region 
[see Eq.~(\ref{eq:LC})]. 
Since the Levi-Civita spacetime is everywhere untrapped, the spacetime 
of interest has an untrapped spacelike asymptotic region. 
Thus, the inequality (\ref{eq:hayward}) implies that the spacetimes that 
we consider are everywhere untrapped. We have
\begin{equation}
\dot{R}^2-{R'}^2=r^2\dot{\beta}^2-(\beta+r\beta')^2\longrightarrow -\beta^2|_{r=0}
~~~{\rm for}~~r\longrightarrow0.
\end{equation}
Because we have $\beta|_{r=0}=e^{\gamma}|_{r=0}>0$, 
the symmetry axis must be untrapped. 
Equation (\ref{eq:hayward}) and the above equation lead to 
\begin{equation}
{R'}^2>\beta^2|_{r=0}+\dot{R}^2. 
\end{equation}
Since $R$ is a regular function, the above inequality implies 
that $R'>0$ everywhere because 
$R'|_{r=0}=\beta|_{r=0}>0$. Therefore, $R$ is positive for $r>0$, 
and this means that $\beta$ is positive in the physical domain. 

\section{Non-Radial Geodesics Near the Singularity} 

In this appendix, we show that non-radial geodesics cannot intersect 
the spacetime singularity at $r=0$. Using Eq.(\ref{eq:c-quantities}), 
the normalization of the geodesic tangent vector is given by
\begin{equation}
e^{-2(\gamma-\psi)}\left[
\left(\frac{dt}{d\tau}\right)^2
-\left(\frac{dr}{d\tau}\right)^2
\right]
+\frac{L^2e^{2\psi}}{r^2\beta^2}
+e^{-2\psi}P^2=\chi,
\end{equation}
where $\chi=-1$ for timelike geodesics, $\chi=1$ for spacelike 
geodesics and $\chi=0$ for null geodesics. Using this normalization 
condition, we rewrite the $r$-components of the geodesic equations 
(\ref{eq:r-geodesic}) in the form 
\begin{equation}
\frac{d}{d\tau}\left(e^{2(\gamma-\psi)}\frac{dr}{d\tau}\right)
=\frac{L^2 e^{2\psi}}{r^3\beta}
+\frac{f(t,r)}{r^2},
\end{equation}
where 
\begin{equation}
f(t,r)=L^2e^{2\psi}\left(\frac{\beta'}{\beta}-\psi'\right) 
+r^2\gamma'\left(\chi-e^{-\psi}P^2\right)
-r^2\psi'\left(\chi-2e^{-\psi}P^2\right).
\end{equation}
Because the metric variables $\beta~(>0)$, $\gamma$ and $\psi$ are regular 
functions of $t$ and $r$ ($\geq0$), there exists a positive 
real number $\varepsilon$ such that 
\begin{equation}
\sup\left\{rf(t,r): 0\leq r <\varepsilon\right\}<1
\end{equation}
holds. Thus, if we consider a positive real number ${\cal C}$ satisfying 
\begin{equation}
{\cal C}<\frac{\sup\left\{ (L\beta e^{\gamma})^2: 0 \leq r < \varepsilon \right\}}
{1-\sup\left\{rf(t,r): 0\leq r <\varepsilon\right\}},
\end{equation}
then we have 
\begin{equation}
\frac{d}{d\tau}\left(e^{2(\gamma-\psi)}\frac{dr}{d\tau}\right)
>\frac{\cal C}{r^3}e^{2(\psi-\gamma)}.
\end{equation}
Because we consider the future directed causal geodesics  
approaching the naked singularity at $r=0$, 
$dr/d\tau$ must be negative. Then, multiplying both sides of the 
above inequality by $e^{2(\gamma-\psi)}dr/d\tau$, we have
\begin{equation}
\frac{d}{d\tau}\left[
\frac{1}{2}e^{4(\gamma-\psi)}\left(\frac{dr}{d\tau}\right)^2
+\frac{\cal C}{r^2}\right]<0
\end{equation}
for $0\leq r<\varepsilon$. From the above equation, there should be a 
positive constant ${\cal H}$ such that
\begin{equation}
\frac{1}{2}e^{4(\gamma-\psi)}\left(\frac{dr}{d\tau}\right)^2
+\frac{\cal C}{r^2}<{\cal H}.
\end{equation}
Thus, we have
\begin{equation}
r>\sqrt{\frac{\cal C}{\cal H}}.
\end{equation}
The above inequality implies that non-radial geodesics cannot intersect the 
naked singularity at $r=0$. 

\section{Christoffel Symbols in the Cartesian Coordinate System} 

The Christoffel symbols in the Cartesian coordinate system are given by 
\begin{eqnarray*}
\Gamma^t_{tt}&=&\dot{\gamma}-\dot{\psi},~~~
\Gamma^t_{tx}=\left(\gamma'-\psi'\right)\cos\varphi,~~~
\Gamma^t_{ty}=\left(\gamma'-\psi'\right)\sin\varphi, \\
\Gamma^t_{xx}&=&\dot{\beta}\beta e^{-2\gamma}\sin^2\varphi
+\dot{\gamma}\cos^2\varphi
-\dot{\psi}\left(\cos^2\varphi+\beta^2e^{-2\gamma}\sin^2\varphi\right), \\
\Gamma^t_{xy}&=&-\left[\dot{\beta}\beta e^{-2\gamma}-\dot{\gamma}
+\dot{\psi}\left(1-\beta^2 e^{-2\gamma}\right)\right]\sin\varphi\cos\varphi, \\
\Gamma^t_{yy}&=&\dot{\beta}\beta e^{-2\gamma}\cos^2\varphi
+\dot{\gamma}\sin^2\varphi-\dot{\psi}\left(\sin^2\varphi
+\beta^2 e^{-2\gamma}\cos^2\varphi\right),\\
\Gamma^t_{zz}&=&\dot{\psi}e^{-2\gamma+4\psi}, \\
\Gamma^x_{tt}&=&(\gamma'-\psi')\cos\varphi, \\ 
\Gamma^x_{tx}&=&\frac{\dot{\beta}}{\beta}\sin^2\varphi
+\dot{\gamma}\cos^2\varphi-\dot{\psi}, ~~~~
\Gamma^x_{ty}=-\left(\frac{\dot{\beta}}{\beta}-\dot{\gamma}\right)
\sin\varphi\cos\varphi, \\
\Gamma^x_{xx}&=&\biggl[\frac{\beta'}{\beta}
\left(2-\beta^2 e^{-2\gamma}\right)\sin^2\varphi+\gamma'\cos^2\varphi
-\psi'\left\{1+\left(1-\beta^2e^{-2\gamma}\right)\sin^2\varphi\right\} 
\nonumber \\
&+&\frac{1}{r}\left(1-\beta^2 e^{-2\gamma}\right)\sin^2\varphi\biggr]
\cos\varphi, \\
\Gamma^x_{xy}&=&\biggl[\frac{\beta'}{\beta}
\left(\beta^2 e^{-2\gamma}\cos^2\varphi-\cos2\varphi\right)
+\gamma'\cos^2\varphi
-\psi'\left(\beta^2e^{-2\gamma}\cos^2\varphi+\sin^2\varphi\right) 
\nonumber \\
&-&\frac{1}{r}\left(1-\beta^2 e^{-2\gamma}\right)\cos^2\varphi\biggr]
\sin\varphi, \\
\Gamma^x_{yy}&=&\biggl[-\frac{\beta'}{\beta}
\left(\beta^2 e^{-2\gamma}\cos^2\varphi+2\sin^2\varphi\right)
+\gamma'\sin^2\varphi
+\psi'\left(\beta^2e^{-2\gamma}\cos^2\varphi+\sin^2\varphi\right) 
\nonumber \\
&+&\frac{1}{r}\left(1-\beta^2 e^{-2\gamma}\right)\cos^2\varphi\biggr]
\cos\varphi, \\
\Gamma^x_{zz}&=&-\psi'e^{-2\gamma+4\psi}\cos\varphi, \\
\Gamma^y_{tt}&=&(\gamma'-\psi')\sin\varphi,\\
\Gamma^y_{tx}&=&-\left(\frac{\dot{\beta}}{\beta}-\dot{\gamma}\right)
\sin\varphi\cos\varphi,~~~~
\Gamma^y_{ty}=\frac{\dot{\beta}}{\beta}\cos^2\varphi
+\dot{\gamma}\sin^2\varphi-\dot{\psi},\\
\Gamma^y_{xx}&=&\biggl[-\frac{\beta'}{\beta}
\left(\beta^2 e^{-2\gamma}\sin^2\varphi+2\cos^2\varphi\right)
+\gamma'\cos^2\varphi
+\psi'\left(\beta^2e^{-2\gamma}\sin^2\varphi+\cos^2\varphi\right) 
\nonumber \\
&+&\frac{1}{r}\left(1-\beta^2 e^{-2\gamma}\right)\sin^2\varphi\biggr]
\sin\varphi, \\
\Gamma^y_{xy}&=&\biggl[\frac{\beta'}{\beta}
\left(\beta^2 e^{-2\gamma}\sin^2\varphi+\cos2\varphi\right)
+\gamma'\sin^2\varphi
-\psi'\left(\beta^2e^{-2\gamma}\sin^2\varphi+\cos^2\varphi\right) 
\nonumber \\
&-&\frac{1}{r}\left(1-\beta^2 e^{-2\gamma}\right)\sin^2\varphi\biggr]
\cos\varphi, \\
\Gamma^y_{yy}&=&\biggl[\frac{\beta'}{\beta}
\left(2-\beta^2 e^{-2\gamma}\right)\cos^2\varphi+\gamma'\sin^2\varphi
-\psi'\left\{1+\left(1-\beta^2e^{-2\gamma}\right)\cos^2\varphi\right\} 
\nonumber \\
&+&\frac{1}{r}\left(1-\beta^2 e^{-2\gamma}\right)\cos^2\varphi\biggr]
\sin\varphi, \\
\Gamma^y_{zz}&=&-\psi'e^{-2\gamma+4\psi}\sin\varphi, \\
\Gamma^z_{tz}&=&\dot{\psi},~~~~
\Gamma^z_{xz}=\psi'\cos\varphi,~~~
\Gamma^z_{xz}=\psi'\sin\varphi.
\end{eqnarray*}
We can easily see that if $\beta'$, $\gamma'$ and $\psi'$ vanish at $r=0$, 
all of the Christoffel symbols are single-valued. 

\section{Riemann Tensor}

The components of the Riemann tensor with respect to the coordinate basis 
are given by
\begin{eqnarray*}
R^{tr}{}_{tr} &=&+e^{2(\psi-\gamma)}
\left(
\ddot{\gamma}-\gamma''-\ddot{\psi}+\psi'' 
\right),
\\
R^{tz}{}_{tz} &=&+e^{2(\psi-\gamma)}
\left(
\ddot{\psi}-\dot{\gamma}\dot{\psi}-\gamma'\psi'+2\dot{\psi}^2+{\psi'}^2
\right),
\\
R^{tz}{}_{rz} &=&+e^{2(\psi-\gamma)}
\left(
\dot{\psi}'-\gamma'\dot{\psi}-\dot{\gamma}\psi'+3\psi'\dot{\psi}
\right),
\\
R^{rz}{}_{rz} &=&-e^{2(\psi-\gamma)}
\left(
\psi''-\dot{\gamma}\dot{\psi}-\gamma'\psi'+\dot{\psi}^2+2{\psi'}^2
\right),
\\
R^{\varphi t}{}_{\varphi t} &=&-e^{2(\psi-\gamma)}
\Biggl[
\frac{1}{r}\left(\gamma'-\psi'\right)+\ddot{\psi}-\dot{\gamma}\dot{\psi}
-\gamma'\psi'+{\psi'}^2+\frac{\dot{\beta}}{\beta}\left(\dot{\gamma}+\dot{\psi}\right)
\nonumber \\
&&~~~~~~~~~+\frac{\beta'}{\beta}\left(\gamma'-\psi'\right)-\frac{\ddot{\beta}}{\beta}
\Biggr],
\\
R^{\varphi t}{}_{\varphi r} &=&-e^{2(\psi-\gamma)}
\Biggl[
\frac{1}{r}\left(\dot{\gamma}-\frac{\dot{\beta}}{\beta}\right)
+\dot{\psi}'-\gamma'\dot{\psi}-\dot{\gamma}\psi'+{\psi'}\dot{\psi}
+\frac{\dot{\beta}}{\beta}\dot{\gamma} 
+\frac{\beta'}{\beta}\gamma'-\frac{\dot{\beta}'}{\beta}
\Biggr],
\\
R^{\varphi r}{}_{\varphi r} &=&+e^{2(\psi-\gamma)}
\Biggl[
\frac{1}{r}\left(\gamma'+\psi'-2\frac{\beta'}{\beta}\right)+\psi''-\dot{\gamma}\dot{\psi}
-\gamma'\psi'+\dot{\psi}^2+\frac{\dot{\beta}}{\beta}\left(\dot{\gamma}-\dot{\psi}\right)
\nonumber \\
&&~~~~~~~~~~+\frac{\beta'}{\beta}\left(\gamma'+\psi'\right)-\frac{\beta''}{\beta}
\Biggr],
\\
R^{\varphi z}{}_{\varphi z} &=&-e^{2(\psi-\gamma)}
\left(\frac{\psi'}{r}+\dot{\psi}^2-{\psi'}^2-\frac{\dot{\beta}}{\beta}\dot{\psi}
+\frac{\beta'}{\beta}\psi'
\right),
\end{eqnarray*}
with all other components vanishing.

\end{document}